%% file: main.tex
\documentclass[useAMS,usenatbib,fleqn]{mnras}

\usepackage{times}
\usepackage{subfigure}
\usepackage{amsmath}
\usepackage{amssymb}
\usepackage{amsbsy}
\usepackage{bmpsize}
\usepackage{graphicx}
\usepackage{paralist}
\usepackage{mathrsfs}
\usepackage{booktabs}
\usepackage{tabularx}
\usepackage{courier}
\usepackage{verbatim}
\usepackage{lscape}
\newcommand{\bs}[1]{\boldsymbol{#1}}
\newcommand{\mat}[1]{\mathbfss{#1}}

\newcommand{\gc}{\mathrm{gc}}
\newcommand{\percent}{\,\mathrm{per\,cent}}

\title[Shapes and alignments of Local Group galaxies]{The shapes and alignments of the satellites of the Milky Way and Andromeda}
\author[J. L. Sanders \& N. Wyn Evans]{Jason L. Sanders\thanks{E-mail: jls@ast.cam.ac.uk} and N. Wyn Evans\thanks{E-mail:nwe@ast.cam.ac.uk}\\
Institute of Astronomy, Madingley Road, Cambridge, CB3 0HA}
\date{Accepted XXX. Received YYY; in original form ZZZ}

\begin{document}
\label{firstpage}
\pagerange{\pageref{firstpage}--\pageref{lastpage}} \pubyear{2016}
\maketitle
\begin{abstract}
We measure the intrinsic shapes and alignments of the dwarf spheroidal (dSph) galaxies of the Local Group. We find the dSphs of the Milky Way are intrinsically flatter (mean intrinsic ellipticity $\mu_E\sim0.6$) than those of M31 ($\mu_E\sim0.5$) and that the classical Milky Way dSphs ($M_V<-8.5\,\mathrm{mag}$) are rounder ($\mu_E\sim0.5$) than the ultrafaints ($\mu_E\sim0.65$) whilst in M31 the shapes of the classical and ultrafaint dSphs are very similar. The M31 dSphs are preferentially radially aligned with a dispersion of $\sim45\deg$. This signal is driven by the ultrafaint population whilst the classical M31 dSphs are consistent with a random orientation. We compare our results to the Aquarius mock stellar catalogues of Lowing et al. and find the subhalo radial alignment distribution matches the Local Group dSphs results, whilst the Aquarius intrinsic ellipticities are significantly smaller than the data ($\Delta\langle E\rangle\approx0.4$). We provide evidence that the major axes of the Milky Way satellites lie within a preferential plane with normal vector pointing towards $(\ell,b)=(127,5)\deg$. We associate this preferred direction with the Vast Polar Orbital structure although their respective great circles are offset by $\sim30\deg$. No signal in the alignments of the major axes is found in M31, suggesting that the Great Plane of Satellites is formed from recent accretion or chance alignment. Finally, we provide predictions for the discrepancy between the velocity dispersion versus scale radius distributions for the Milky Way and M31 populations and demonstrate that the projection effect from viewing similar populations from two different locations does not account for the discrepancy which is probably caused by increased tidal disruption in M31. 
\end{abstract}


\section{Introduction}
The $\Lambda$ cold dark matter (CDM) cosmological paradigm predicts that density fluctuations in the early Universe were magnified by the accretion of dark matter and these early dark matter haloes provided potential wells in which baryons could cool and begin forming galaxies. The subsequent evolution of the Universe was hierarchical with small dark matter haloes accreting onto ever larger haloes. A natural manifestation of this picture is the presence of a spectrum of dark matter subhaloes within each halo \citep{Diemand2007,Springel2008} which are believed to host the dwarf spheroidal (dSph) galaxies of the Local Group (LG).  The dSphs are heavily dark-matter dominated environments \citep{Mateo1998,Kleyna2001,McConnachie2012} and so provide ideal laboratories to test the $\Lambda$CDM paradigm -- as the less well understood effects of baryons are sub-dominant in these systems. Recent years have seen the number of known dSphs in the Milky Way (MW) and Andromeda (M31) increase significantly (approximately 60 spectroscopically confirmed dSphs fainter than $M_V=-13.5\,\mathrm{mag}$ are known) such that studies based on the statistical properties of the dSph population are possible. In particular, matching the shape and structure of the dSph population is a key test of the $\Lambda$CDM picture.

From a theoretical perspective, dark matter haloes are generically triaxial \citep[e.g.][]{JingSuto2002,BailinSteinmetz2005,Allgood2006}. Both field haloes and subhaloes are typically more flattened and more triaxial with increasing mass \citep{Allgood2006}, and subhaloes tend to be rounder than field haloes due to tidal effects \citep{Kuhlen2007,VeraCiro2014}. For MW-mass haloes, the addition of baryons makes the haloes rounder~\citep{Debattista2008}, but it appears that at lower subhalo mass scales, baryons have very little effect on shape of the dark matter haloes \citep{Knebe2010}. 

A further prediction from dark-matter-only cosmological simulations is that dark matter subhaloes tend to have major axes that align with the vector towards the centre of their host halo \citep{PereiraBryan,Kuhlen2007,Faltenbacher2008,Barber2015}. Filamentary accretion produces a weak alignment signal for haloes outside the virial radius and once subhaloes pass inside the virial radius, tidal interaction increases the degree of radial alignment independent of the subhalo mass \citep{PereiraBryan}. After approximately one orbit, the initially triaxial subhaloes become tidally locked to the host halo and so spend a large fraction of their orbits pointing towards the halo centre (around pericentre, the subhalo cannot respond fast enough to remain aligned). Additional figure rotation against the orbital direction delays the tidal locking of the subhaloes \citep{PereiraBryan,Barber2015}.

However, it appears from several studies that galaxies in clusters are essentially consistent with random alignment \citep[e.g.][]{Schneider2013} and so it is acknowledged that dark matter only simulations may produce `too much' radial alignment \citep{Kiessling2015,Joachimi2015}. A solution to this discrepancy uses more centrally bound dark-matter particles as proxies for baryons for which the degree of alignment is weaker \citep{PereiraBryan,Knebe2008b} as tidal locking takes $\sim$ twice as long for a more centrally-concentrated stellar population \citep{PereiraBryan2010}. However, other baryonic effects are important on the cluster scale and the degree of radial alignment from full hydrodynamical simulations on these scales is an on-going area of active research \citep{Tenneti2015a}, particularly for interpretation of weak cosmological gravitational lensing signals \citep{Kiessling2015,Joachimi2015}. The effect of baryons on the Milky Way subhalo scale is still an open question with, for instance, \cite{Knebe2010} suggesting baryons have little effect on the alignment of the dark matter haloes at these low mass scales.

\begin{figure}
\includegraphics[width=\columnwidth]{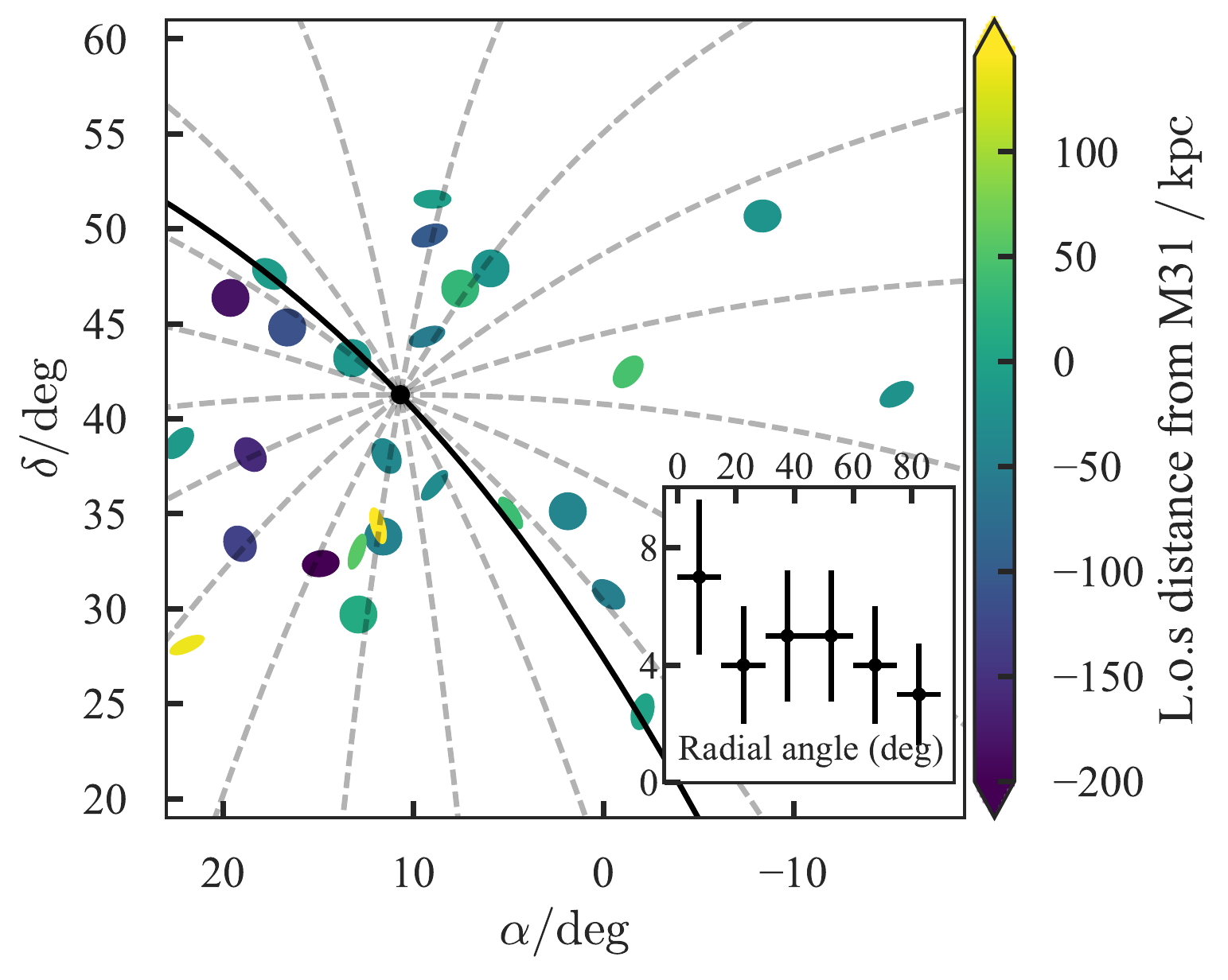}
\caption{Shape and alignment of the M31 dSphs. Each dSph is depicted as an ellipse showing the axis ratio and on-sky orientation (in a local Cartesian basis). The colour shows the distance along the line of sight from the plane in which M31 lies. The black line shows the M31 disc plane and the dashed grey lines show great circles through the centre of M31 (shown by a black dot)\protect\footnotemark.}
\label{Fig::M31}
\end{figure}

Observations of the Local Group provide an alternative route to constraining the properties of dark matter on subhalo scales. For instance, we show the on-sky distribution of the M31 dSphs in Figure~\ref{Fig::M31} along with the distribution of misalignments between the on-sky major axis and the on-sky radial direction. As already noted by, e.g., \cite{Barber2015}, the distribution of radial misalignment angles is not uniform suggesting there is evidence for preferential radial alignment in the Local Group. Indeed, \cite{Salomon2015} has already used the M31 dataset to measure the intrinsic shapes and alignments, assuming all dSphs are prolate figures. These authors in general find a mean intrinsic ellipticity of $\sim0.5$ and that both a purely tangentially-aligned model and a randomly-aligned model are preferred over a purely radially-aligned model, as the latter fails to reproduce the roundest dSphs.

There is already some evidence that the dSph population of the Local Group is unusual when compared to simulations. In both the Milky Way and Andromeda, the dSphs seem to be preferentially grouped in planes. \cite{LyndenBell1976} first noted that in the Milky Way the dSph galaxies seem to lie along a great circle that also coincides with the Magellanic stream and this picture has been reinforced in recent years by the discovery of more dSphs that may also reside in the plane \cite[the so-called Vast Polar Orbital structure, VPOS,][]{Pawlowski2015}. Additionally, \cite{Ibata2013} discovered a similar plane of Andromeda dSph galaxies, the so-called Great Plane of Satellites (GPOS). It is therefore quite natural to ask whether the shapes and alignments of the dSphs support these interpretations of the data. 

\footnotetext{The projection employed is not conformal so the angles depicted between the major axes and the great circles (in the main panel) are not exactly the true angles (shown in the inset).}

In this paper, we measure the intrinsic distribution of shapes and alignments of the dwarf spheroidal galaxies of the Local Group. The central idea behind our work is that if the MW and M31 host similar populations of dSphs, then we can exploit the two distinct viewing perspectives to extract the intrinsic shape and alignment distributions. With the MW dSphs alone, we have little leverage on measuring the radial alignment distribution, and the MW dSphs can break some of the degeneracies in extracting the intrinsic shape distribution using just the M31 dSphs. 

We begin by presenting the framework for converting between intrinsic and observed properties of an optically thin galaxy stratified on self-similar triaxial ellipsoids in Section~\ref{Sec::Framework}. In Section~\ref{Sec::Analysis}, we present three models for the intrinsic shape and alignment of the dSph population and discuss the results of fitting these models to the data. In Section~\ref{Sec::Simulations}, we compare the results of our analysis to the tagged Aquarius simulations provided by \cite{Lowing2015} and in Section~\ref{Sec::VelDisp}, we discuss how the predictions for the velocity dispersion against scale radius relation from our models. In Section~\ref{Sec::Conclusions}, we present our conclusions.

\section{Framework}\label{Sec::Framework}

\begin{figure}
\includegraphics[width=\columnwidth]{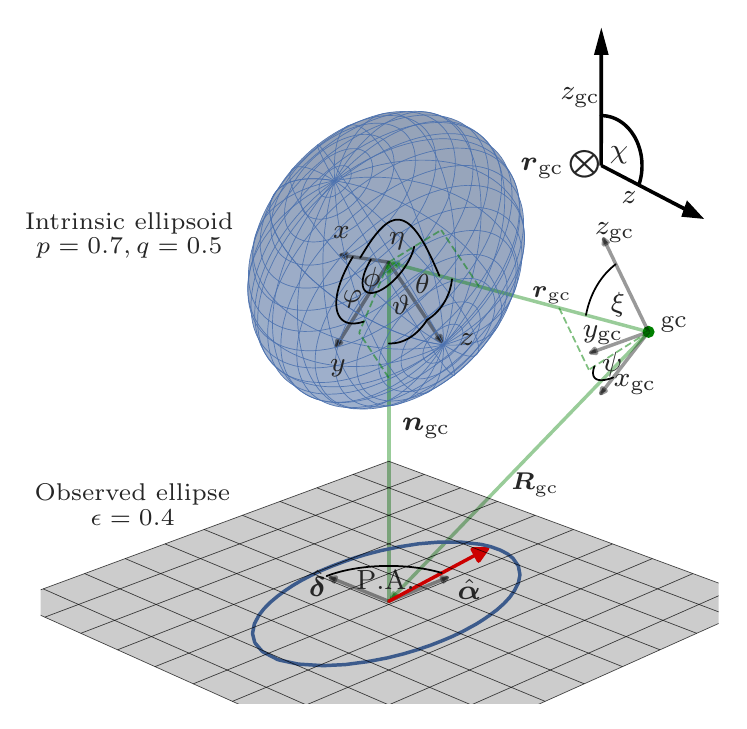}
\caption{Dwarf spheroidal coordinate definition: the dSph is located at the displacement vector $\bs{r}_\gc$ from the centre of its host galaxy (gc) and displacement vector $\bs{n}_\gc$ from the observer, who sits at $\bs{R}_\gc$ relative to the galaxy centre. An equidensity contour of the dSph is shown by the blue mesh. In the intrinsic spherical polar coordinate system of the dSph $(x,y,z)$, the vector $-\boldsymbol{r}_\mathrm{gc}$ is at angles $(\theta,\phi)$ and $-\boldsymbol{n}_\mathrm{gc}$ is at $(\vartheta,\varphi)$. The angle between the major axis and the radial vector is given by $\eta$. The observer sees the projected ellipse shown at the bottom of the diagram with its major axis aligned at a position angle of $\mathrm{P.A.}$ with respect to North through East.}
\label{Fig::Diagram}
\end{figure}

In this section, we give the transformations between the orientation of a dSph with respect to the centre of its host galaxy and the properties of the dSph observed from Earth. Throughout this paper, we assume that dSphs are well modelled as optically-thin and the light distribution is stratified on self-similar ellipsoids. We also assume the dSphs are sufficiently small and distant (i.e. $r_h/D\ll1$) such that the observed shape can be computed assuming the galaxies are observed from infinity. Fig.~\ref{Fig::Diagram} gives a diagrammatic definition of many of the quantities used in this section.

\subsection{Ellipsoidal systems}

We begin with a brief primer on the properties of ellipsoids and their projected ellipses \citep{Contopoulos1956,Binney1985,Franx1991,Evans2000}. An ellipsoid is characterised by two axis ratios: the intermediate-to-major axis ratio $p=b/a$ and the minor-to-major axis ratio $q=c/a$. It is convenient to define the intrinsic ellipticity $E$ and the triaxiality $T$:
\begin{equation}
\begin{split}
E&=1-q>0,\\
T&=(1-p^2)/(1-q^2).
\end{split}
\end{equation}
An oblate spheroid has $T=0$, whilst a prolate spheroid has $T=1$.

If an ellipsoid is viewed by an observer located at infinity, it appears as an ellipse. The viewing angles $\varphi$ and $\vartheta$  are defined as standard spherical polar angles with respect to the principal axes $(\bs{\hat{x}},\bs{\hat{y}},\bs{\hat{z}})$ of the dSph (in the order major, intermediate, minor) i.e. $\tan\phi=|\bs{\hat{y}}|/|\bs{\hat{x}}|$ and $\cos\theta=|\bs{\hat{z}}|$. The ellipticity of the observed ellipse, $\epsilon(p,q,\varphi,\vartheta)$, is given by
\begin{equation}
(1-\epsilon)^2 = \frac{A-\sqrt{B}}{A+\sqrt{B}},
\label{Eqn::Ellipticity}
\end{equation}
where
\begin{equation}
\begin{split}
A=(&1-q^2)\cos^2\vartheta-(1-p^2)\sin^2\vartheta\sin^2\phi+p^2+q^2,\\
B=(&(1-q^2)\cos^2\vartheta-(1-p^2)\sin^2\vartheta\sin^2\phi-p^2+q^2)^2+\\&4(1-p^2)(1-q^2)\sin^2\vartheta\cos^2\vartheta\sin^2\varphi.
\end{split}
\end{equation}
The observed minor axis of the ellipse is offset from the projected ellipsoidal minor axis by an angle $f(p,q,\varphi,\vartheta)$ given by
\begin{equation}
\tan2f = \frac{2T\sin\varphi\cos\varphi\cos\vartheta}{\sin\vartheta-T(\cos^2\varphi-\sin^2\varphi\cos^2\vartheta)}.
\label{Eqn::MinorAxisAngle}
\end{equation}

\subsection{Conversion between observed and galactocentric properties}
We work in a Galactocentric right-handed Cartesian coordinate system where $\bs{\hat{x}}_\gc$ points from the centre of the Milky Way towards the Sun, $\bs{\hat{y}}_\gc$ points in the opposite direction to Galactic rotation and $\bs{\hat{z}}_\gc$ points towards the North Galactic pole. We introduce the displacement vector $\bs{R}_\gc$ between the host galactic centre and the Sun, and the displacement vector $\bs{r}_\gc$ between the centre of the dSph and the host galaxy. The displacement vector between the Sun and the dSph is then given by $\bs{n}_\gc=\bs{r}_\gc-\bs{R}_\gc$.

An observer at the centre of the host galaxy observes the dSph at viewing angles $(\theta,\phi)$. There is a third degree of freedom described by the angle $\chi$ which gives the rotation of the on-sky basis vectors for an observer at the centre of the galaxy i.e. the direction of North ($\chi$ is formally the right-handed rotation angle about the galactocentric radial vector $\bs{r}_\gc$). For an observer situated at the galactic centre, $\chi$ only affects the zero-point of the rotation angle of the observed major axis and so does not change the observed properties of the dSph. However, it does affect the viewing angle of an observer at some other location -- we can think of $\chi$ as rotating the entire galaxy about a vector pointing towards a fixed dSph.

We wish to transform coordinates in the Galactocentric Cartesian basis into the intrinsic basis of the dSph such that given a general vector $\bs{n}_\mathrm{gc}$, we can compute the viewing angles $(\vartheta,\varphi)$ and the orientation of the projected ellipsoidal minor axis on the sky with respect to equatorial North, $\mathrm{P.A.}_\mathrm{proj}$. Note that the orientation of the projected ellipsoidal minor axis is not the same as the minor axis of the observed ellipse and they are related via equation~\eqref{Eqn::MinorAxisAngle}.

We introduce the Euler rotation matrix $\mat{R}(\alpha,\beta,\gamma) = \mat{R}_z(\alpha)\mat{R}_x(\beta)\mat{R}_z(\gamma)$ where
\begin{equation}
\mat{R}_z(a) = 
\begin{pmatrix}
\cos a&-\sin a&0\\
\sin a&\cos a&0\\
0&0&1
\end{pmatrix},
\mat{R}_x(b) = 
\begin{pmatrix}
1&0&0\\
0&\cos b&-\sin b\\
0&\sin b&\cos b
\end{pmatrix}.
\end{equation}

We first perform a rotation to align the radial vector $\bs{r}_\gc$ with respect to the $z''$ axis of an on-sky coordinate system $(x'',y'',z'')$ for an observer at the galactic centre. Setting $\xi=\arccos z_\gc/r_\gc$ and $\psi=\arctan y_\gc/x_\gc$, the corresponding rotation matrix is given by $\mat{R}_{\mathrm{gc,obs}}=\mat{R}(0,\xi,\pi/2-\psi)$. Note this aligns the $y''$ axis with $\bs{\hat\xi}$. Next we perform the rotation $\mat{R}_{\mathrm{obs,int}}=\mat{R}(\phi+\pi/2,\theta-\pi,-\chi)$ between the `galactocentric observer' frame and the intrinsic dSph frame ($\chi$ gives the angle between the projected intrinsic minor axis and the galactic North pole for an observer at the galactic centre. If $\chi=0$ the minor axis of the dSph lies in the $x''=0$ plane). This rotates to the intrinsic coordinate system of the dSph such that $\bs{r}_\gc= -|\bs{r}_\gc|(\sin\theta\cos\phi,\sin\theta\sin\phi,\cos\theta)$ (in the intrinsic basis of the dSph). The combined matrix $\mat{R}_{\mathrm{obs,int}}\mat{R}_{\mathrm{gc,obs}}$ rotates any vector $\bs{n}$ from galactocentric coordinates ($\bs{x}_\gc$) to the intrinsic coordinate system ($\bs{x}$) such that $\vartheta=\arccos n_{z}/|\bs{n}|$ and $\varphi=\arctan n_{y}/n_{x}$.

A final consideration is the observed on-sky orientation of the dSph. Given $(\vartheta,\varphi)$, equation~\eqref{Eqn::MinorAxisAngle} gives the angle between the observed minor axis and the projected ellipsoidal minor axis, so we therefore must compute the orientation of the projected ellipsoidal minor axis for the observer. We first compute the minor axis direction in the galactocentric frame as $\mat{R}_{\mathrm{gc,obs}}^{-1}\mat{R}_{\mathrm{obs,int}}^{-1}\bs{\hat{z}}$ (where $\bs{\hat{z}}$ is expressed in the dSph basis as $\bs{\hat{z}}=(0,0,1)$). We then rotate this vector to galactic coordinates using $\mat{R}_{\mathrm{gc,gal}}=\mat{R}(0,b-\pi/2,\pi/2-l)$. A further rotation takes us to equatorial coordinates which enables us to compute the angle of the projected minor axis with respect to equatorial North, $\mathrm{P.A.}_\mathrm{proj}$. Note that this is distinct from the observed minor axis from equation~\eqref{Eqn::MinorAxisAngle}. The observed position angle of the major axis is given by
\begin{equation}
\mathrm{P.A.}=\mathrm{P.A.}_\mathrm{proj}+f+\pi/2.
\label{Eqn::PArelation}
\end{equation}

The angle, $\eta$, between the galactocentric radial vector and the major axis is given by
\begin{equation}
\cos\eta = \sin\theta\cos\phi.
\end{equation}

A sanity check of our procedure is to draw uniform samples in $\bs{r}_\gc$, $\chi$, $\phi$ and $\cos\theta$, and check the distributions of $\cos\vartheta$, $\varphi$ and $\mathrm{P.A.}_\mathrm{proj}$ are also uniform and display no correlations with $\ell$ and $b$. Additionally, placing the observer at a large distance and sampling uniformly in $\chi$ and $\phi$ produces a uniform distribution in $\cos\vartheta$ irrespective of the distribution of $\theta$.

\begin{table}
\caption{Data used in this study: ellipticities ($\epsilon$), position angles of major axes (P.A.) and distance moduli $\mu$. The first section gives the data for the classical MW dSphs (brighter than $M_V=-8.5\,\mathrm{mag}$), the second the ultrafaints of the MW, the third the classical M31 dSphs and the final section the ultrafaint M31 dSphs.}
\begin{minipage}{\columnwidth}
\begin{tabular}{lccc}
\input{data_latex.dat}
\end{tabular}
\end{minipage}
\label{Table::Data}
\end{table}

\section{Shape and alignment inference}\label{Sec::Analysis}
For our data, we use all confirmed dwarf spheroidal galaxies of the Milky Way and Andromeda with absolute V-band magnitude $>-13.5\,\mathrm{mag}$ (additionally we remove Leo T due to its high gas content). There are 33 Milky Way dwarf spheroidals and 28 Andromeda dwarf spheroidals in our sample with properties listed in Table~\ref{Table::Data}. Our data are taken primarily from the updated tables of \cite{McConnachie2012} complemented by the recent analysis of the M31 dSphs by \cite{Martin2016} in addition to data from \cite{Torrealba2016a,Torrealba2016b}. We also opt to analyse just the `classical' dSphs which are defined as the nine Milky Way dwarfs brighter than $M_V=-8.5\,\mathrm{mag}$ (Canes Venatici I, Carina, Draco, Fornax, Leo I, Leo II, Sextans, Sculptor, Ursa Minor) and equivalently the 19 Andromeda dSphs brighter than $M_V=-8.5\,\mathrm{mag}$. 

For our modelling, we use PyStan \citep[][a Python wrapper for the Stan probabilistic programming language]{stan,pystan} to construct probabilistic models for the intrinsic distribution of shapes and alignments. In this section, we describe each of the models employed. Using PyStan, we sample from our models using a Hamiltonian Monte Carlo algorithm (the in-built NUTS sampler \cite{NUTSsampler}). Four separate chains are run and we ensure the Gelman-Rubin convergence diagnostic is close to unity ($\hat{R}<1.005$) for each of the parameters. The results of our model fits are given in Table~\ref{Table::Results}. We will now discuss the specifics of the modelling and discuss the results. 

\begin{table*}
\caption{Results table: medians and 1-sigma intervals for the parameters in our model fits (maximum marginalized posterior and likelihoods given in brackets below each entry -- these are identical for location parameters for which we use uniform priors). The three sets of rows are for the random alignment model, the radially aligned model and the planar-aligned model. Each set of rows is broken down into the Milky Way and M31 subsets which in turn are divided into the classical (C) ($M_V<-8.5\,\mathrm{mag}$) and ultrafaint (UF) subsets. The first four parameters describe the mean $\mu$ and standard deviation $\sigma$ of the triaxiality $T$ and ellipticity $E$ distributions. $\sigma_{\cos\eta}$ gives the intrinsic spread about radial alignment. $\ell_n$ and $b_n$ give the Galactic angles of the preferred plane with the intrinsic spread of major axes about this as $\sigma_{\sin\alpha}$. }
\input{all_results.dat}
\label{Table::Results}
\end{table*}

\subsection{Asymmetric uncertainties}\label{Sec::AsymmErrors}

Our primary data are the ellipticities and position angles given in Table~\ref{Table::Data}, which often come with asymmetric error bars. As we lack the specific pdfs of the measured quantities, we must approximate their posteriors. We use the `linear-variance' approximation from \cite{Barlow2004} such that for a measured value of the ellipticity of $\epsilon^{+\sigma_{\epsilon+}}_{-\sigma_{\epsilon-}}$ the likelihood on the ellipticity is
\begin{equation}
p(\epsilon'|\epsilon,\sigma_{\epsilon+},\sigma_{\epsilon-})\propto\mathcal{G}(\epsilon'-\epsilon,\sigma_{\epsilon+}\sigma_{\epsilon-}+(\sigma_{\epsilon+}-\sigma_{\epsilon-})(\epsilon'-\epsilon)),
\label{Eqn::Barlow}
\end{equation}
where $\mathcal{G}(\mu,\sigma^2)$ is a Gaussian with mean $\mu$ and variance $\sigma^2$. This distribution has the problem that the variance can be negative so we truncate the variance at $1\times10^{-4}$. For a sample of M31 dSphs in the PAndAS survey, \cite{Martin2016} has supplied posterior samples for the ellipticity and position angle from fits to the photometry, which we can use to check our approximate posteriors. We have found that for small asymmetric errors, equation~\eqref{Eqn::Barlow} matches the posterior distributions well but when $\epsilon-\sigma_{\epsilon-}<0$ (i.e. the ellipticity is consistent with zero), the distributions are better approximated as
\begin{equation}
p(\epsilon'|\epsilon,\sigma_{\epsilon+},\sigma_{\epsilon-})\propto\mathcal{G}(\epsilon',\sigma_{\epsilon+}^2+\sigma_{\epsilon-}^2),
\label{Eqn::ZeroEps}
\end{equation}
Any $95\percent$ upper bound measurements $\epsilon_u$ are treated as measurements of zero with an uncertainty of $\epsilon_u/2$. We use the same linear-variance approach for the position angle and distance modulus measurements with no further adjustment.

In Appendix~\ref{Appendix}, we discuss the validity of the approximation used in equation~\eqref{Eqn::Barlow} by comparison to models which explicitly employ the posterior samples from \cite{Martin2016}. We find that the posterior distributions on the model parameters are very similar when using the posterior samples compared to the approximation employed throughout the paper. Additionally, in the absence of similar sets of posterior samples for the other M31 dSphs and all of the MW dSphs we are forced to employ an approximation.

We now detail each of the models employed. The first model assumes the dSphs are randomly aligned with respect to the observer, so does not utilize the position angle information. The second model assumes the major axis of the dSphs are preferentially aligned with the galactocentric radial vector. Our final model assumes the major axes of the dSphs preferentially lie in a plane.

\subsection{Random alignment}
We attempt to infer the distribution of $T$ and $E$ given the observed shapes $\epsilon$ and associated uncertainties under the assumption that there is no preferred orientation for the dSphs. Implicitly, this assumes the distribution of observed position angles is uniform. Our model is described as
\begin{equation}
\begin{split}
\mu_T,\mu_E&\sim\mathcal{U}(0,1),\\
\sigma_T,\sigma_E&\sim\mathcal{C}(0,1),\\
\cos\vartheta_i&\sim\mathcal{U}(0,1),\\
\varphi_i&\sim\mathcal{U}(0,\tfrac{1}{2}\pi),\\
T_i&\sim\mathcal{N}(\mu_T,\sigma_T),\\
E_i&\sim\mathcal{N}(\mu_E,\sigma_E),\\
\epsilon_i&\sim\mathcal{N}'(\epsilon(\cos\vartheta_i,\varphi_i,T_i,E_i),\sigma_{\epsilon+i},\sigma_{\epsilon-i}),
\end{split}
\label{Eqn::RandomModel}
\end{equation}
where $\mathcal{N}'$ is the modified normal distribution to account for the asymmetric uncertainties, and subscript $i$ corresponds to the $i$th of $N_\star$ dSphs. $\mathcal{U}(a,b)$ is a uniform distribution between $a$ and $b$ and $\mathcal{C}(a,b)$ is a Cauchy distribution centred on $a$ with scale parameter $b$.  From dark-matter-only simulations, the triaxiality and ellipticity distribution are well approximated by Gaussians \citep{Allgood2006,Knebe2008a}. The natural choice for the prior on a scale parameter $\sigma$ would be a Jeffrey's prior of $1/\sigma$. Instead, we heed the advice of \cite{Gelman2006} and adopt a half-Cauchy distribution as a weakly informative prior. Very large values of $\sigma$ are disfavoured, whilst very small values are not strongly favoured. This matches nicely our expectation that the intrinsic triaxiality and ellipticity distributions will not be $\delta$-function spikes nor will they lack any structure at all. The choice of the scale as unity is inspired by the finite range of both the triaxiality and the ellipticity.

One complication is that some of our variables have a discrete range so the pdf that describe them must be normalized over this range. In the case of the uniformly distributed variables, the normalization is a constant factor but for the normally distributed variables (i.e. $T_i$ and $E_i$) the normalization depends on the parameters $\mu$ and $\sigma$ that we are attempting to constrain. Therefore, any distributions $f(x)$ for a variable $x$ constrained to lie in the interval $(a,b)$ that are quoted in this paper implicitly contain a factor of $N^{-1}$ where
\begin{equation}
N(f(x),a,b) = \int_a^b \mathrm{d}x\,f(x).
\end{equation}
We additionally introduce the pdf $\mathcal{G'}$ given by
\begin{equation}
\mathcal{G'}(\mu,\sigma^2) = \frac{\mathcal{G}(\mu,\sigma^2)}{N(\mathcal{G}(\mu,\sigma^2),0,1)}.
\end{equation}

For clarity, the log-likelihood for this model is given by
\begin{equation}
\begin{split}
\ln\mathcal{L} = &\sum_i^{N_\star}\ln \int\mathrm{d}\epsilon'\,\mathrm{d}\cos\theta\,\mathrm{d}\phi\,\mathrm{d}T\,\mathrm{d}E\, p(\epsilon_i|\epsilon',\sigma_{\epsilon_+},\sigma_{\epsilon_-})\\&\times p(\epsilon'|\cos\theta,\phi,T,E)\mathcal{G}'(T|\mu_T,\sigma_T^2)\mathcal{G}'(E|\mu_E,\sigma_E^2),
\end{split}
\label{Eqn::LogLikelihood}
\end{equation}
where $p(\epsilon'|\cos\theta,\phi,T,E)$ is a $\delta$-function related to equation~\eqref{Eqn::Ellipticity}.
\begin{figure}
\includegraphics[width=\columnwidth]{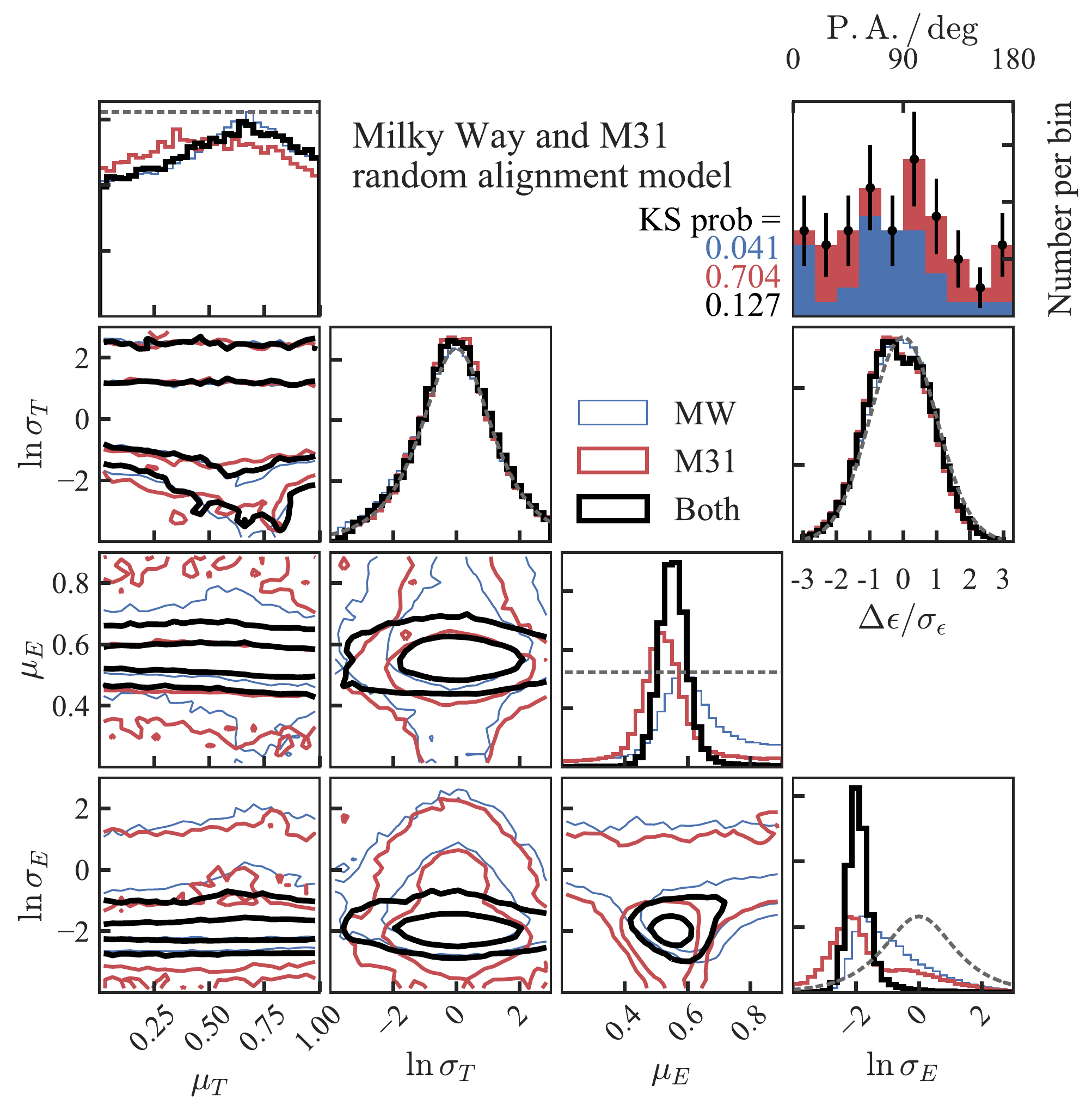}
\caption{Inference for the random alignment model. The corner plot gives the 2d correlations between the model parameters (we show contours containing $68\percent$ and $95\percent$ of the probability) and the 1d distributions on the diagonal (the dashed grey lines show the adopted prior distributions). The shape distributions are modelled as Gaussian distributions described by the mean $\mu$ and variance $\sigma^2$ in the triaxiality $T$ and ellipticity $E$. The thin blue contours show the results for just the Milky Way dSphs, the thicker red for just the M31 dSphs and the thick black for both samples combined. The top extra panel shows the distribution of position angles decomposed into MW (blue) and M31 (red) dSphs along with the Kolmogorov-Smirnov probabilities that these data are drawn from a uniform distribution. The bottom extra panel shows the difference between the samples of the ellipticity and the ellipticity data normalized by the reported uncertainties.}
\label{Fig::RandomResults}
\end{figure}

\subsubsection{Results}
The posterior distributions for the model parameters are shown in Fig.~\ref{Fig::RandomResults} and the corresponding medians and $1\sigma$ intervals are provided in Table~\ref{Table::Results}. We also provide maximum marginalized posterior and likelihood values (identical for the location parameters for which we use uniform priors). When these two quantities differ significantly, the results are highly prior-dependent and uncertain. We see that the triaxiality distribution parameters are poorly measured with the width parameter $\sigma_T$ following the prior and all allowed values of $\mu_T$ being near equally probable (there is a weak bias using the MW data for more prolate models i.e. higher $T$). \cite{Binggeli1980} showed that inferences on the triaxiality distribution were poor using only photometric data. The ellipticity parameters on the other hand are much better measured. We find that the median $\mu_E$ for the Milky Way ($\mu_E=0.61^{+0.20}_{-0.12}$) is slightly larger than that for M31 ($\mu_E=0.52^{+0.10}_{-0.09}$) such that the dSphs of M31 are in general slightly rounder. However, the spread parameter $\sigma_E$ is less well constrained for the Milky Way, highlighting there is a larger spread about this median flattest dSph. The combination of MW and M31 provides a measurement of $\mu_E = 0.55\pm0.05$.

In Table~\ref{Table::Results}, we give a further breakdown into the classical and ultrafaint dSphs in the MW and M31. We find that in the MW the classical dSphs are intrinsically rounder ($\mu_E=0.49^{+0.24}_{-0.16}$) than the ultrafaint dSphs ($\mu_E=0.65^{+0.22}_{-0.31}$) and have a larger spread of intrinsic ellipticities ($\sigma_E\approx0.8$ for the ultrafaints compared to $\sigma_E\approx0.4$ for the classical dSphs). This was already observed by \cite{Martin2008} using a sample of 15 dSphs. However, in M31 we do not find a significant difference between the mean intrinsic ellipticities of the classical and ultrafaint dSphs, although as with the MW the classical dSphs have a narrower spread in the intrinsic ellipticity distribution. When comparing the MW to M31, we find that the classical dSph populations are very similar, whilst the ultrafaint populations are more distinct with the M31 ultrafaints being generally rounder. This may be a reflection of the fainter magnitudes (and hence masses) reached for the MW dSphs. Additionally, the width of the intrinsic ellipticity distribution for all the dSphs in the MW appears to be narrower ($\sigma_E=0.31$) than that in M31 ($\sigma_E=0.16$). It is worth noting that dark-matter-only simulations at the Milky Way scale \citep{Kuhlen2007} and cluster scale \citep{Knebe2008a} find that subhaloes become flatter with increasing mass.

Our results can be directly compared to those of \cite{SanchezJanssen}, who also fitted the triaxiality and ellipticity distributions to the MW and M31 dSphs assuming no preferred viewing angle. \cite{SanchezJanssen} used 23 Local Group dSphs brighter than $M_V=-8\,\mathrm{mag}$, so we can compare to our results using the classical dSphs. These authors find $\mu_E=0.51^{+0.07}_{-0.06}$ and $\sigma_E=0.12^{+0.06}_{-0.05}$ exactly in agreement with our result of $\mu_E=0.5\pm0.05$ and $\sigma_E=0.10^{+0.06}_{-0.04}$. Interestingly, these authors also find that the mean triaxiality is constrained ($\mu_T\approx0.5$) which we do not find, but this may due to the use of different priors. 

In Fig.~\ref{Fig::RandomResults}, we show the residuals between the model and data ellipticities indicating the model is a good fit\footnote{Here we have taken $\sigma_\epsilon=\sqrt{\sigma_{\epsilon+}^2+\sigma_{\epsilon-}^2}/\sqrt{2}$. As the ellipticity distributions are not Gaussian, the residuals do not perfectly match a Gaussian distribution. In particular, our procedure breaks down for the ellipticity measurements of zero so, for plotting purposes only, we randomly make half the residuals negative so the residual distribution better approximates a Gaussian.}. We also show the distribution of position angles and the associated Kolmogorov-Smirnov probability that these distributions are drawn from a uniform distribution (i.e. that expected from this random model). These indicate there is weak evidence that the position angles are in tension with the random orientation model, so we now go on to investigate two further models.

\subsection{Preferred alignment}
We now explore whether the data favour a particular orientation for the dSphs. As discussed in the introduction, cosmological simulations tend to produce subhaloes that have their major axes radially aligned with their host halo. Our data are now supplemented with position angles $\mathrm{P.A.}^{+\sigma_{\mathrm{P.A.}+i}}_{-\sigma_{\mathrm{P.A.}-i}}$ and positions $(\ell,b,\mu\pm\sigma_\mu)$ where we assume no uncertainty on the Galactic coordinates $(\ell,b)$ and use the Gaussian uncertainty on the distance modulus $\mu$ (given in Table~\ref{Table::Data}). From these measurements, we can compute $\bs{n}_\mathrm{gc}$ and combined with $\bs{R}_\mathrm{gc}$ we can find $\bs{r}_\mathrm{gc}$ (from Section~\ref{Sec::Framework}).

We expand the model of the previous section as
\begin{equation}
\begin{split}
\cos\theta_i&\sim\mathcal{U}(-1,1),\\
\phi_i&\sim\mathcal{U}(-\tfrac{\pi}{2},\tfrac{\pi}{2}),\\
\chi_i&\sim\mathcal{U}(0,\pi),\\
\sigma_{\cos\eta}&\sim\mathcal{C}(0,1),\\
|\cos\eta_i|&\sim\mathcal{N}(1,\sigma_{\cos\eta}),\\
\mathrm{P.A.}_i&\sim\mathcal{N}'(P(\bs{r}_\mathrm{gc},\phi,\theta,\chi),\sigma_{\mathrm{P.A}+i},\sigma_{\mathrm{P.A}-i}),
\end{split}
\label{Eqn::AlignModel}
\end{equation}
where we have only listed the changes to the model presented in equation~\eqref{Eqn::RandomModel}. $\vartheta$ and $\varphi$ are now deterministic variables and the function $P$ is given by $P=\mathrm{P.A.}_\mathrm{proj}+f+\pi/2$ (equation~\eqref{Eqn::PArelation}). Note the position angle is constrained to $0<\mathrm{P.A.}<\pi$ and if $|P-\mathrm{P.A}|>\pi/2$ we take $P+\pi$ if $P>\mathrm{P.A.}$ and $P-\pi$ otherwise. The chosen model is a Gaussian in the cosine of the misalignment angle between the radial vector and major axis ($\eta$). \citet{Knebe2008a} has found in dark-matter-only simulations that a $\cos^4\eta$ distribution matches the alignment distribution over all mass scales probed. Here, we adopt a normal distribution for its simplicity.

\subsubsection{Results}
\begin{figure}
\includegraphics[width=\columnwidth]{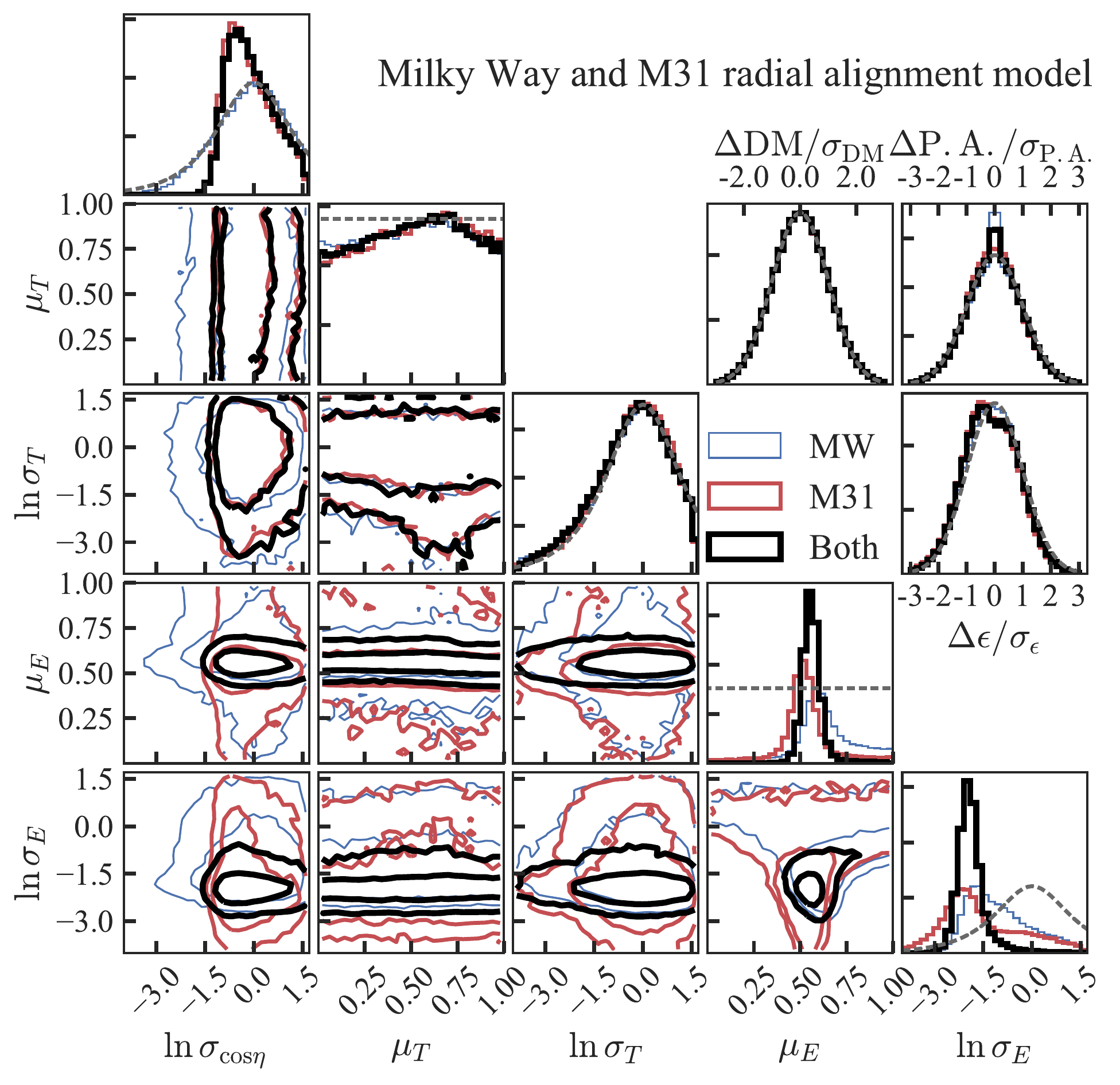}
\caption{Inference for the radial alignment model. The corner plot gives the 2d correlations between the model parameters (we show contours containing $68\percent$ and $95\percent$ of the probability) and the 1d distributions on the diagonal (the dashed grey lines show the adopted prior distributions). The parameter $\sigma_{\cos\eta}$ gives the intrinsic scatter about pure radial alignment $\eta=0$ and the shape parameters are as in the random alignment case. The thin blue contours show the results for just the Milky Way dSphs, the thicker red for just the M31 dSphs and the thick black for both samples combined. The other three panels show the distributions of the sample with respect to the measured values normalized by the uncertainties for the distance modulus, position angle and ellipticity (from top left to bottom right).}
\label{Fig::AlignResults}
\end{figure}

The results of fitting this model to the data are shown in Fig.~\ref{Fig::AlignResults} and given in Table~\ref{Table::Results}. We also show the residuals for the ellipticities, position angles and distance moduli. Note the peak in position angle is produced by the unconstrained position angles (Crater II and Horologium II). We see that the posterior distributions for the shape parameters are very similar to the results obtained in the random alignment case. When considering solely the MW dSphs, we find that there is no constraint on the spread about radial alignment $\sigma_{\cos\eta}$ as the posterior exactly follows the prior. This is exactly per expectation where, as we are observing very near the radial direction for the dSphs, it is very difficult to measure how radially aligned the dSphs are. However, for M31 we observe the population from a different perspective and we find there is weak evidence for preferential radial alignment in the M31 dSph population with $\sigma_{\cos\eta}\sim0.45$ (maximum likelihood) corresponding to an angular spread of $\sim45\deg$ although there is a long tail to much larger values. We note though that the posterior distribution is much peakier than the prior so the signal is genuine and matches expectation from Fig.~\ref{Fig::M31}. When combining the MW and M31 dSphs we find very similar conclusions to just the M31 case. It appears that given the current data quality we do not gain any significant leverage on the shape and alignment distributions of the dSph population by using two perspectives.

\begin{figure}
\includegraphics[width=\columnwidth]{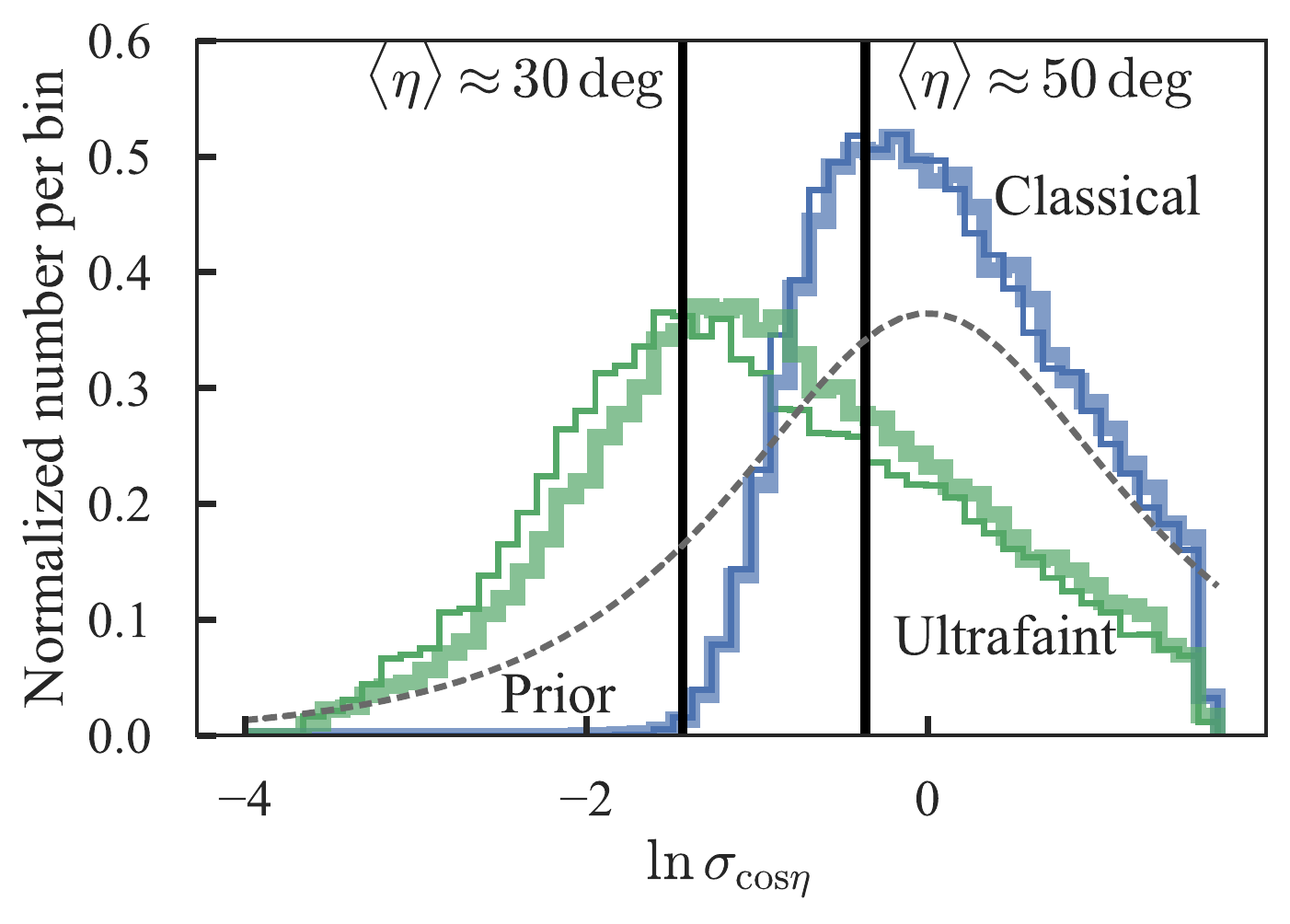}
\caption{Posterior distributions for the spread about radial alignment $\sigma_{\cos\eta}$. The thick lines show the results using just the M31 dSphs whilst the thin lines show the addition of the MW dSphs. The grey dashed line shows the adopted prior. The two vertical lines give the maximum marginalized posterior values for the combined datasets labelled by the approximate angular spread about radial alignment.}
\label{Fig::AlignCUF}
\end{figure}

When we break down the M31 dSphs into classical and ultrafaint groups, we find that the classical M31 dSph population is consistent with being randomly distributed whilst it is the ultrafaint population that has a significant radial alignment signal with $\sigma_{\cos\eta}\approx0.15$ (maximum likelihood) corresponding to an angular spread of $\sim25\deg$. The posterior distributions for $\sigma_{\cos\eta}$ are shown in Fig.~\ref{Fig::AlignCUF}. When considering all the ultrafaint dSphs, we find a maximum posterior value of $\sigma_{\cos\eta}\approx0.24$ ($\sim33\deg$) but a maximum likelihood value of $\sigma_{\cos\eta}\approx0.12$ ($\sim23\deg$), whilst for all the classical dSphs we find $\sigma_{\cos\eta}\approx0.6$ ($\sim50\deg$). This result is in contradiction to dark-matter-only simulations of subhaloes where the degree of radial alignment is observed to be independent of mass \citep{Knebe2008a,Knebe2008b,PereiraBryan}.

\subsection{Planar alignment}
We finally investigate whether the major axes of the dSphs are aligned in a plane. We introduce the vector $\hat{\boldsymbol{n}}=(\cos b_n\cos\ell_n,\cos b_n\sin\ell_n,\sin b_n)$ in the Galactocentric basis and express our new probabilistic model as
\begin{equation}
\begin{split}
\ell_n&\sim\mathcal{U}(0,\pi),\\
\sin b_n&\sim\mathcal{U}(-1,1),\\
\sin\alpha_i&\sim\mathcal{N}(0,\sigma_{\sin\alpha}),\\
\sigma_{\sin\alpha}&\sim\mathcal{C}(0,1),
\end{split}
\end{equation}
where again we have only listed additions and note that there is no constraint on $\eta$ in this model. $\alpha_i$ is the angle between the plane defined by the normal vector $\hat{\boldsymbol{n}}$ and the major axis of the dSph so is given by
\begin{equation}
\sin\alpha_i = \hat{\boldsymbol{n}}\cdot\mat{R}_{\mathrm{gc,obs}}^{-1}\mat{R}_{\mathrm{obs,int}}^{-1}\bs{\hat{x}},
\end{equation}
where $\bs{\hat{x}}$ is expressed in the dSph basis as $\bs{\hat{x}}=(1,0,0)$. The parameters of this model do not depend on the distance modulus to each dSph.

\begin{figure*}
\includegraphics[width=\textwidth]{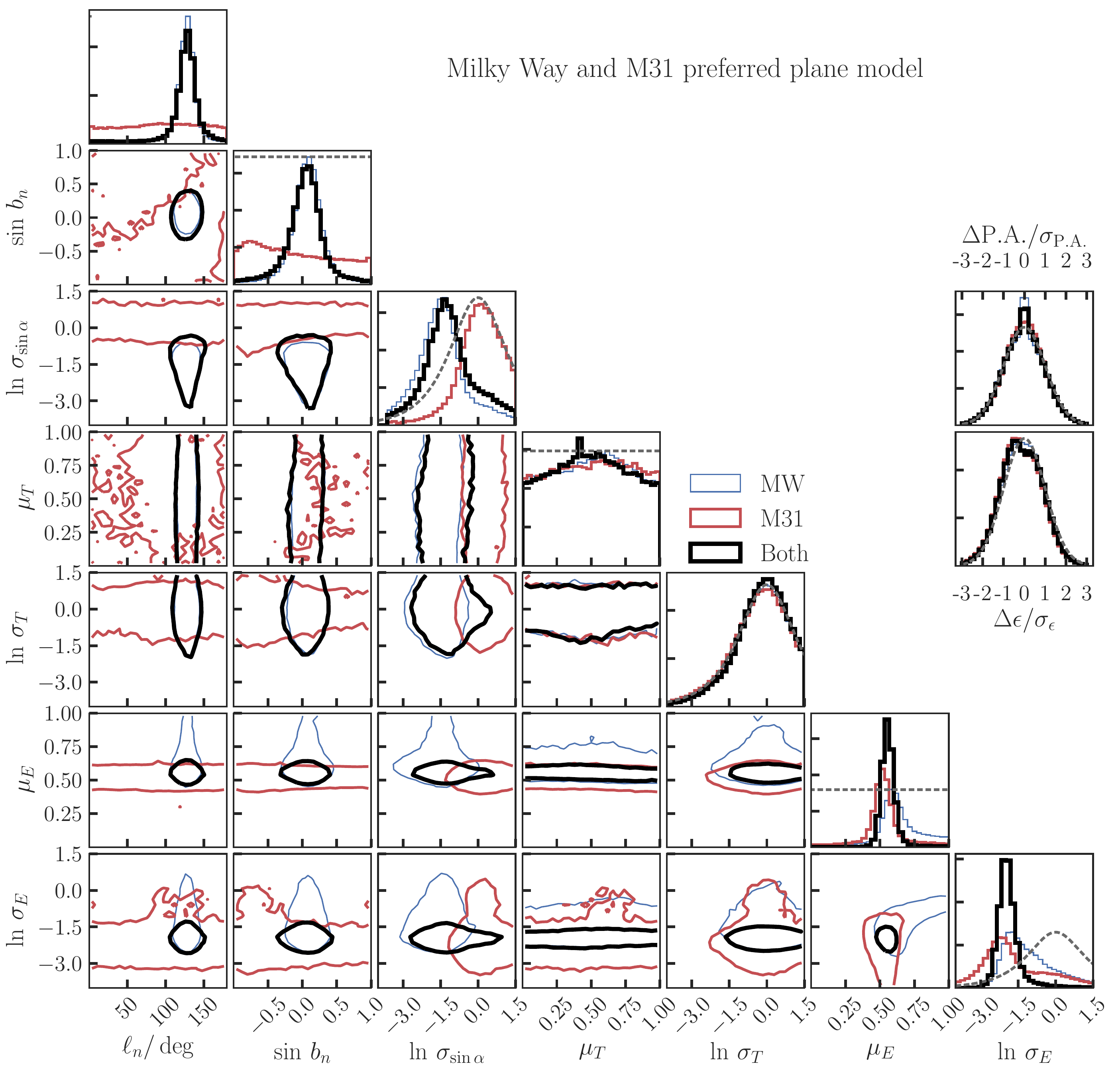}
\caption{Inference for the preferred plane model. The corner plot gives the 2d correlations between the model parameters (we show contours containing $68\percent$ of the probability) and the 1d distributions on the diagonal (the dashed grey lines show the adopted prior distributions). The parameters $\ell_n$ and $\sin b_n$ are the Galactic coordinates that describe the normal to the plane and $\sigma_{\sin\alpha}$ gives the intrinsic scatter about the plane. The four shape parameters are as in the random model case. The thin blue contours show the results for just the Milky Way dSphs, the thicker red for just the M31 dSphs and the thick black for both samples combined. The other two panels show the distributions of the sample with respect to the measured values normalized by the uncertainties for the position angle (top) and ellipticity (bottom).}
\label{Fig::PlaneResults}
\end{figure*}

\subsubsection{Results}
The posterior distributions for the parameters of this model are shown in Fig.~\ref{Fig::PlaneResults}. We see from this figure, as well as from Table~\ref{Table::Results}, that the results for the shape distributions are very similar to the random case. Interestingly, the Milky Way fits produce a very clear peak in the posterior distribution for the direction of the unit vector describing the plane with median $\ell_n=128\deg$ and $b_n=5\deg$. The median spread about this plane is $\sigma_{\sin\alpha}=0.23$ corresponding to an angular spread of $\sim14\deg$. We note that the distribution of the spread parameter $\sigma_{\sin\alpha}$ is significantly different to the prior, indicating that the plane is significant. When including all the M31 dwarfs, the direction of $\hat{\bs{n}}$ moves to slightly lower latitudes and the spread about the plane becomes slightly larger $\sim20\deg$. This indicates that the M31 dSph orientations are not consistent with lying within any particular plane. Indeed, when using just the M31 dSphs, we find no strong evidence for a preferred plane -- there is a small peak in the location parameters at $\ell\approx90\deg$ and $b\approx-50\deg$ but the distribution of the spread parameter $\sigma_{\sin\alpha}$ is much less peaked than in the MW only case and essentially follows the prior indicating that there is no significant plane. This conclusion is in tension with claims of the Great Plane of Satellites (GPOS) in Andromeda \citep{Ibata2013}. However, it seems that although the plane of satellites appears as an overdensity along a great circle through the centre of Andromeda, the members of this plane do not have their major axes preferentially aligned with this plane (this can be seen approximately in Fig.~\ref{Fig::M31}). This echoes the conclusions of \cite{Collins2015} who found that the sizes, masses, luminosities, and metallicities of on- and off-plane M31 satellites are essentially indistinguishable, and the conclusions of \cite{Salomon2015} who found that there is no difference in the morphology and alignment of the on- and off-plane satellites. These results point towards a chance alignment \citep{Gillet2015}. However, a different interpretation is that the GPOS is formed from \emph{recent} group infall \citep{Bowden2014,Angus2016}. If the satellites that form the GPOS have only recently been accreted, tidal locking of the major axis into the orbital plane may not have taken effect. This interpretation is also supported by the theoretical arguments of \citet{Bowden2013} who showed that planes of satellites in the outer parts of triaxial dark haloes cannot generically persist over long times without thickening. However, it is a challenge to explain why the GPOS is thinner than expected if produced by dwarf galaxy associations similar to those observed at the current epoch \citep{Metz2009}.

\begin{figure*}
\includegraphics[width=\textwidth]{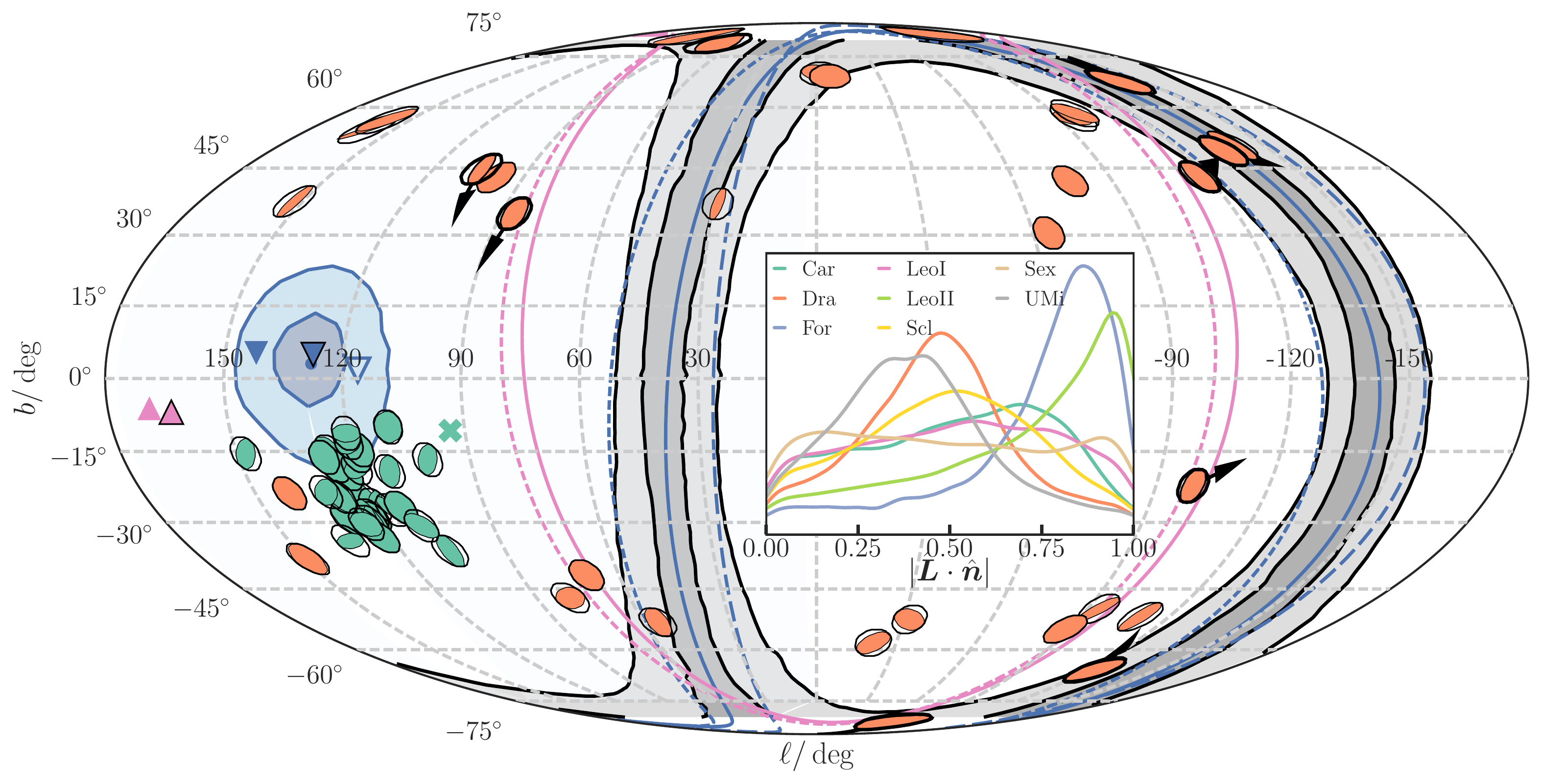}
\includegraphics[width=\textwidth]{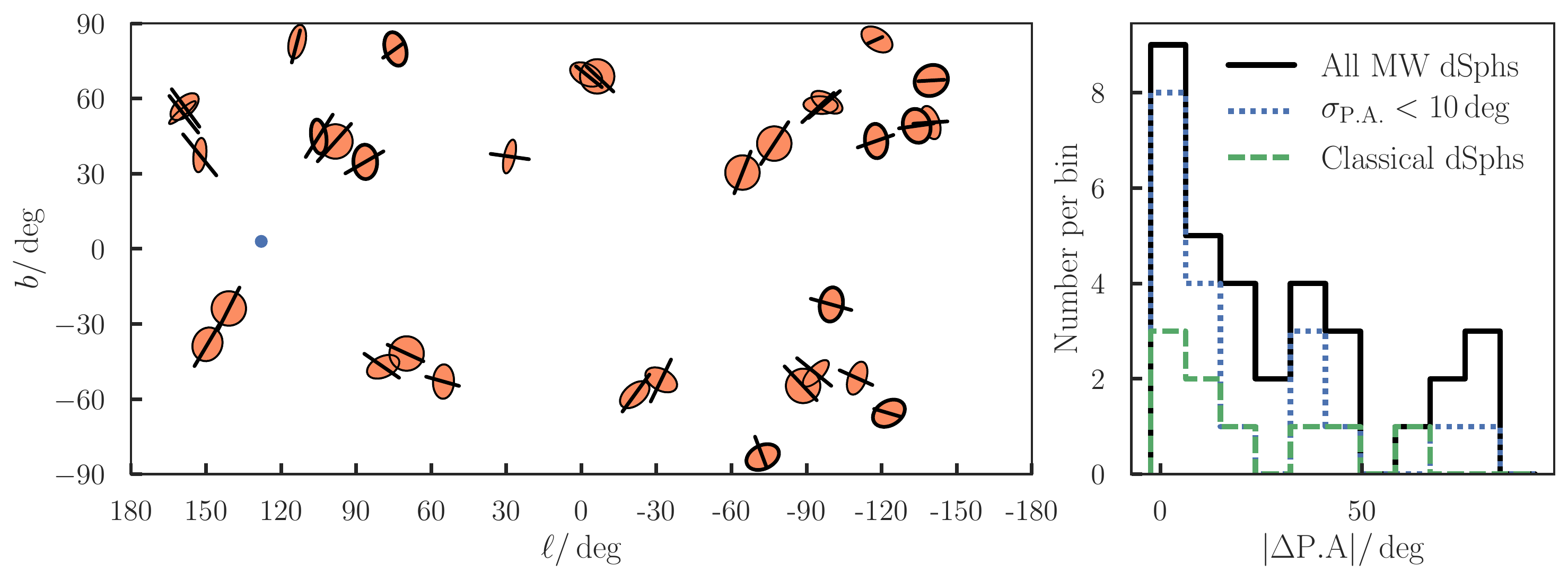}
\caption{Orientation of the discovered plane in Mollweide projection. \textbf{Main panel}: The blue contours show the $32$ and $68\percent$ confidence regions for the location of the normal vector to the preferred plane and the grey contours show the corresponding confidence regions for the intercept of this plane with the celestial sphere. The dSphs of the Milky Way are shown in orange (with the 9 classical dSphs given a thicker black outline) and those of M31 in green. Each dSph is depicted as an ellipse showing the on-sky orientation and observed ellipticity (as these are deformed by the projection we also show a circle at each location in black). The blue line shows the median plane fitted solely to the Milky Way dSphs with normal vector given by the upside-down black-outlined triangle. Using solely the 9 classical MW dSphs, we find the plane shown by the blue dashed line and the normal vector given by the non-outlined upside-down triangle, whilst using the MW ultrafaints we find the plane as the long-dashed blue line with normal vector given by the empty blue triangle. The median normal vector of the plane fitted to M31 is given by the green cross (although the uncertainties in this case are large). The VPOS from \protect\cite{Pawlowski2015} is shown by the solid pink line and its normal vector is given by the outlined triangle (the dashed line and non-outlined triangle show the plane and pole with 4 outliers removed -- VPOS+new-4). \textbf{Inset panel}: For the 9 classical MW dSphs we also show the proper motion vectors (corrected for the Solar motion) and the inset panel shows the posterior distributions for the cosine of the angle between the angular momentum and the normal vector for the plane. \textbf{Lower panels}: The left lower panel shows the dSphs on a Cartesian representation of the Galactic coordinate system. Each dSph is shown as an ellipse in the local on-sky Cartesian coordinates $(\Delta\ell\cos b,\Delta b)$ and the black line segments show the intercept of the great circle that passes through the median normal vector (shown by a blue dot) and the location of each dSph. The right panel shows the distribution of the absolute differences in position angle between the minor axis of each dSph and the corresponding great circle split by those with accurate position angle measurements ($\sigma_\mathrm{P.A.}<10\,\mathrm{deg}$ in short-dashed blue) and just the classical MW dSphs (in long-dashed green).}
\label{Fig::PlaneOnSky}
\end{figure*}

We now investigate further the discovered plane for the Milky Way satellites. In Fig.~\ref{Fig::PlaneOnSky}, we show the on-sky distribution of the MW and M31 dSphs indicating their orientations and flattenings with ellipses. The posterior distribution for the direction of the normal to the plane is indicated along with the posterior for the intercept of the plane with the celestial sphere. We see that the plane is oriented essentially normal to the Galactic disc and passes through Sculptor and Fornax in the Galactic South and Leo I and II in the Galactic North. We also show the claimed Vast Polar Orbital structure (VPOS) plane from \cite{Pawlowski2015} fitted to the spatial distribution of the Milky Way dwarf galaxies (including the Magellanic clouds and the Sagittarius dwarf galaxy). Although our plane does not match exactly with the VPOS (offset by $\sim30\deg$), it is similar in that it is also polar. 

The ellipses in the top panel of Fig.~\ref{Fig::PlaneOnSky} are deformed by the projection employed so, for clarity, the lower panel shows the shapes of the dSphs in their local Cartesian basis set along with segments of great circles that pass through the location of the normal vector $\hat{\bs{n}}$ and the centres of the dSphs. It is clear that the great circle segments preferentially lie near the minor axes of the dSphs, which can be quantitatively seen in the lower right panel  of Fig.~\ref{Fig::PlaneOnSky} where we display the distributions of the position angle difference between the great circle segments and the minor axes $|\Delta\mathrm{P.A.}|$. Both the classical dSphs and the full sample peak near $|\Delta\mathrm{P.A.}|=0$ and the peak becomes more pronounced when restricting ourselves to well measured position angles $\sigma_{\mathrm{P.A.}}<10\,\mathrm{deg}$. The equatorial coordinates of the normal vector are $(\alpha,\delta)\approx(23,67)\,\mathrm{deg}$ which essentially means there is a preference for the dSphs (which are all at declinations less than $67\,\mathrm{deg}$) to have their major axes along lines of constant declination. This can be seen clearly in the upper right panel of Fig.~\ref{Fig::RandomResults} where there is a peak in the MW dSphs at $\mathrm{P.A.}=90\,\mathrm{deg}$. This anisotropy in position angles was noted for the classical dSphs by \cite{LyndenBell1994} but it is striking that it has persisted with the discovery of many fainter satellites.

Our interpretation for the discovered plane is that it is approximately the orbital plane for the satellites. For an orbitting satellite, \cite{Barber2015} has demonstrated that the major axis is more significantly aligned with the galactocentric radial direction than the instantaneous orbital velocity. However, it is reasonable to suppose that tidal torques cause the major axis to lie within the orbital plane of the satellite such that from the centre of the galaxy the major axis \emph{does} appear to be aligned with the orbital direction. The orientation of the major axes of some dSphs was used to lend credence to suggested streams of satellites from \cite{LyndenBell1995}. Our discovered anisotropy of the position angles supports the picture that the satellites are orbitting within the plane in which they currently reside.

One piece of evidence supporting this interpretation of the VPOS is there are a number of satellites with proper motions (corrected for the peculiar velocity of the Sun) that lie within the plane (including the Magellanic clouds). In Fig.~\ref{Fig::PlaneOnSky}, we also show the on-sky solar-motion-corrected velocities for those dSphs with proper motion measurements \citep[][Carina, Draco, Fornax, Leo I, Leo II, Sculptor, Sextans, Ursa Minor]{Pawlowski2013}. As noted by \cite{Pawlowski2013}, there is weak evidence for circulation of the classical dSphs about the polar structure. To quantify this, we have computed the dot product between the normalized angular momentum vector and the normal vector to our discovered plane for a series of samples from the posterior distribution for our joint model fits and from the velocity error ellipsoids. If the plane is long-lived, we anticipate that $|\hat{\bs{L}}\cdot\hat{\bs{n}}|$ peaks at unity. We see that both Fornax and Leo II have peaks in $|\hat{\bs{L}}\cdot\hat{\bs{n}}|$ near unity, Draco and Ursa Minor have a peak near $|\hat{\bs{L}}\cdot\hat{\bs{n}}|\sim0.5$ and all other classical dSphs have much flatter distributions of $|\hat{\bs{L}}\cdot\hat{\bs{n}}|$. This suggests that the discovered plane is not long lived. However, the proper motion measurements are not accurate enough to decide the matter conclusively for the moment. 

We have investigated what happens when fitting planes to samples of $\sim30$ randomly-distributed satellites (both distributed over the whole sky and only at high latitudes $|b|>25\,\mathrm{deg}$). In general, we find a peak in the posterior for the $(\ell_n,b_n)$ distributions which is less significant than that presented here. Additionally, for these checks we find that the posterior distribution of $\sigma_{\sin\alpha}$ approximately follows the prior. This suggests that the plane discovered here is significant and real, but the strength of the signal is perhaps in part due to the selection effects governing the discovery of the dSphs and the corresponding absence of dSphs near the Galactic disc plane. However, we note that the identification of a plane based on the alignments is significantly less affected by selection effects than any identification based on spatial location.

\section{Cosmological simulation predictions}\label{Sec::Simulations}
\begin{figure*}
$$\includegraphics[width=\textwidth]{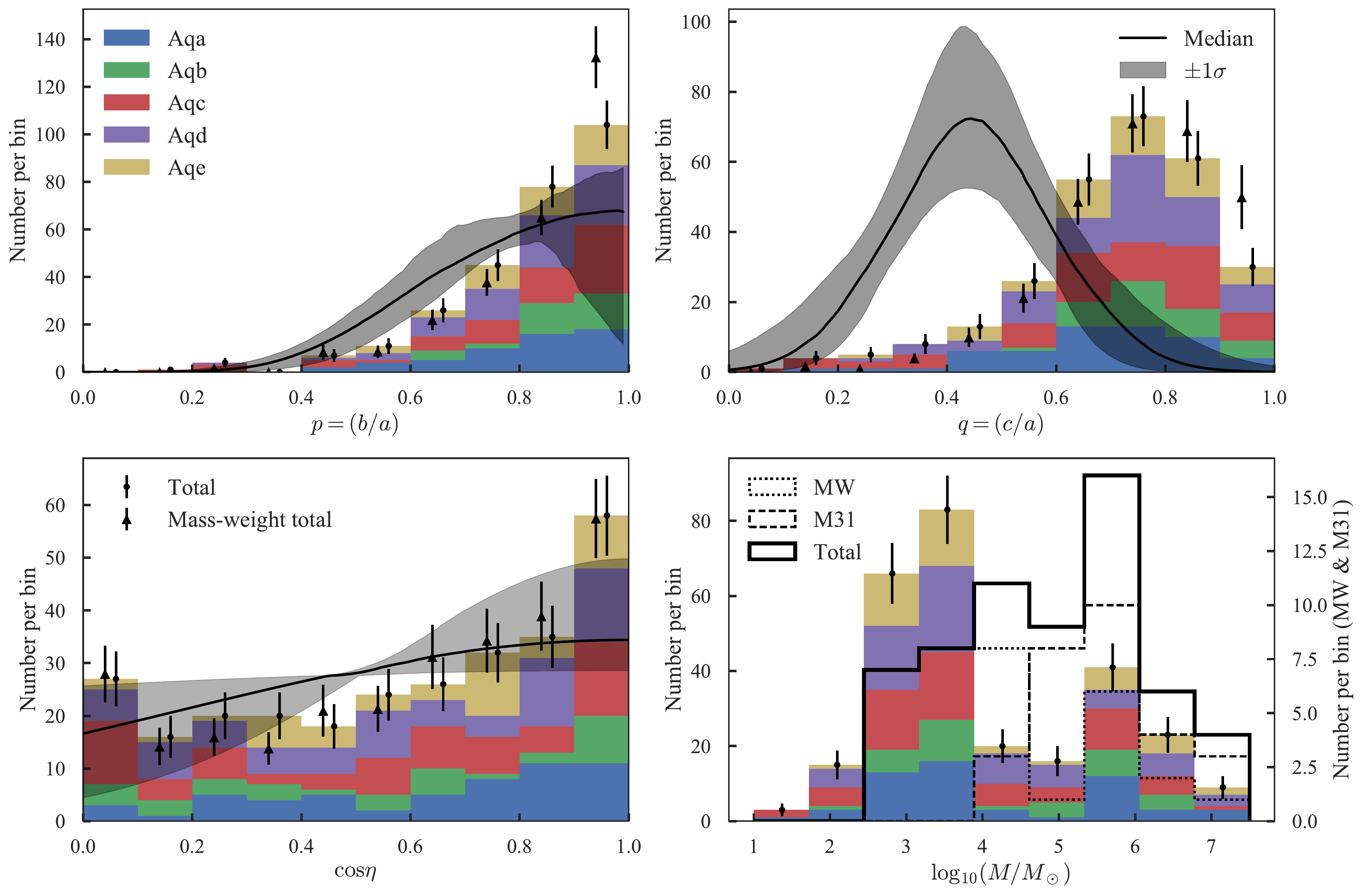}$$
\caption{Properties of the stellar components of the subhaloes from the models of \protect\cite{Lowing2015} using five haloes from the Aquarius simulation series. The coloured bars show the number counts from the five simulations with the black error bars giving the total counts. The black curves and associated grey band are the median and $\pm1\sigma$ region for the radially-aligned model fitted to both the MW and M31 dwarf spheroidals. The first three panels show the intermediate-to-major axis ratio $b/a$, the minor-to-major axis ratio $c/a$ and the cosine of the angle between the major axis and the direction from the centre of the dSph to the centre of the host halo. The final panel shows the mass distribution of the subhaloes. The black line shows the distribution of the whole dSph sample with the short (long) dashed line showing just the MW (M31) dSphs.}
\label{Fig::Aquarius}
\end{figure*}

For our aligned model fitted to both the MW and M31 data, we found a maximum likelihood value of $\sigma_{\cos\eta}=0.48$ and a median of $\sigma_{\cos\eta}=0.84$. This can be directly compared to results from other authors using dark-matter-only simulations. \cite{Kuhlen2007} find that at $r/r_t=0.1$ ($r_t$ is the tidal radius) the median alignment is $\langle\cos\eta\rangle\sim0.55$ which corresponds approximately to $\sigma_{\cos\eta}=1$ but the degree of alignment increases to $\langle\cos\eta\rangle\sim0.75$  and $\sigma_{\cos\eta}=0.37$ at $r/r_t=1$. Similarly, \cite{Faltenbacher2008} find that for subhaloes within the virial radius $\langle\cos\eta\rangle\sim0.55$ whilst \cite{VeraCiro2014} find that $\langle\cos\eta\rangle\sim0.75$ for subhaloes within a radius containing $95\percent$ of the mass. \cite{PereiraBryan} find that for subhaloes within the virial radius the alignment is mass-independent with a median alignment $\langle\cos\eta\rangle\sim0.66$  and $\sigma_{\cos\eta}=0.55$. Despite these simulations being dark-matter-only, it appears that they match our results well, which points towards dark matter being only weakly influenced by and approximately tracing the baryons on the Local Group subhalo scale as suggested by \cite{Knebe2010}.

We now turn to comparing the results of our modelling to expectation from the Aquarius simulations \citep{Springel2008}. These are a series of six Milky Way sized dark matter haloes (each of total mass $\sim10^{12}M_\odot$) chosen randomly from a large cosmological simulation and re-simulated at a series of resolution levels. As such, they are perfect for studying the dark matter subhaloes of the Milky Way. The halo of M31 is believed to be a very similar mass to that of the MW~\citep{Diaz2014} so the Aquarius simulations are also appropriate for the study of M31. \cite{Lowing2015} used the second highest resolution level (maximum particle mass $10^4h^{-1}M_\odot$) to create an accreted stellar halo for five of the Aquarius simulations (AqF was not used as it has two major mergers at $z\sim0.6$). Using a semi-analytic galaxy formation code \citep{Font2011}, the one per cent most bound dark matter particles at each time step are tagged with a total stellar mass and metallicity. At the final simulation snapshot, each tagged particle is then converted into a full stellar population using a set of isochrones and the set of stars is distributed in phase-space over the volume occupied by the parent dark-matter particle. \cite{Lowing2015} provides the resulting stellar catalogues complete with labels indicating the dark-matter subhalo to which the stars are still bound.

These catalogues are ideal for simulating a realistic population of dSph galaxies. However, they naturally have a number of limitations \citep{Cooper2010,Bailin2014,LeBret2015,Cooper2016}. As the baryons are added by tagging the dark-matter particles, the stellar distributions will naturally match the dark matter distribution (albeit the most one per cent most bound) rather than potentially following a more flattened distribution or even living on completely different orbits. Additionally, the addition of baryons does not affect the distribution of dark matter. It is expected that due to the high mass-to-light ratio in dSphs the effect of the baryons on the dark matter is small \citep{Knebe2010}. Finally, the simulations do not include a disc which is anticipated to have a significant effect on the lifetime and structure of subhaloes \citep{DOnghia2010}.

From the \citet{Lowing2015} catalogues, we extract all the stars associated with subhaloes (Subhalo ID$>0$) that lie within the virial radius of the host halo \cite[taken from][]{Springel2008}. The shapes and alignments of the subhaloes are computed by diagonalizing the reduced inertia tensor, $I_{ij}$:
\begin{equation}
I_{ij} = \frac{\sum_n m_n (x_{i,n}-\langle x_i\rangle)(x_{j,n}-\langle x_j\rangle)/d_n^2}{\sum_n m_n/d_n^2},
\end{equation}
where $n$ indexes the stars and the weighting factor of $d_n^2=x_{n}^2+(y_n/p)^2+(z_n/q)^2$ is introduced to down-weight the importance of particles at large radii \citep{Gerhard1983}. The axis ratios are determined by the square-root of the eigenvalue ratios and the radial alignment is found as $\cos\eta = \hat{\bs{e}}_1\cdot\langle\bs{x}\rangle$. As $I_{ij}$ depends on the axis ratios, we compute $I_{ij}$ iteratively by first setting $p=q=1$, computing $p$, $q$ and the eigenvectors, realigning the coordinate system and repeating until $p$ and $q$ both differ by no more than two per cent. The distributions of the axis ratios, the radial alignment and the stellar masses are shown in Fig.~\ref{Fig::Aquarius}. We also show the distributions of these quantities from our radially-aligned model of the Milky Way and M31. From $1000$ posterior samples, we generate the median and $\pm1\sigma$ distributions of $p$, $q$ and $\cos\eta$. The distributions of $q$ and $\cos\eta$ are trivially given by our model, whilst the calculation of the distribution of $p$ for each set of parameters is given by
\begin{equation}
\begin{split}
p(p|\mu_T,&\sigma_T,\mu_E,\sigma_E) = \int_{1-b/a}^1\mathrm{d}E\,\Big|\frac{\partial T}{\partial p}\Big|_E\\\times&\mathcal{G}'(T(p,E)-\mu_T,\sigma_T^2)\mathcal{G}'(E-\mu_E,\sigma_E^2).
\end{split}
\end{equation}
Additionally, we show the stellar mass distributions of our sample of MW and M31 dSphs computed from the absolute V-band magnitude assuming a mass-to-light ratio of unity. The mass distributions for the data and simulations are quite distinct as many of the lower mass subhaloes are too faint to be seen (or the efficiency of star formation at low masses is incorrect in the simulations). We therefore opt to approximate the selection function of the known dSphs by the ratio of the mass distributions and we show the $p$, $q$ and $\cos\eta$ distributions weighted by this `selection function'. Clearly, the selection effects are quite weak\footnote{We also experimented with computing luminosity-weighted reduced inertia tensors and weighting the distributions by the luminosity distribution ratio. The conclusions are very similar although the simulation results are noisier as we are more dominated by very bright single stars in the low-mass systems such that some computations of the reduced inertia tensor for low mass haloes do not converge.}.

The shape distributions from the Aquarius subhaloes do not match the corresponding distributions from the models exactly. The intermediate-to-major axis ratio $p=(b/a)$ distribution is broader in the models than in the simulations, indicating that the centres of the Aquarius haloes are more oblate than the dSph population appears. Additionally, the mass-weighted distribution, although broadly very similar to the non-mass-weighted distribution, has a larger peak near $b/a\sim1$ showing that the more massive subhaloes in Aquarius are slightly more oblate than the lower mass subhaloes \citep[][finds that $b/a$ measured at the location of maximum circular velocity decreases with increasing mass for field haloes but the decrease is much more gradual for the subhaloes]{VeraCiro2014}. It is perhaps surprising that the data appear more triaxial than the simulations when baryons tend to form more axisymmetric components. However, as we have seen, the results on triaxiality are not strong and highly prior-dependent. The minor-to-major axis ratio distributions are clearly in disagreement with the Aquarius haloes, which are significantly rounder than the models (the peaks differ by $\Delta(c/a)\sim0.4$). The conclusion is not affected by mass weighting. If the shape of low-mass dark matter subhaloes are unaffected by any central baryonic component \citep[e.g.][]{Knebe2010} then we conclude that the stellar populations in dSphs are intrinsically flatter than their dark matter haloes by approximately a factor of $2$. However, further hydrodynamic simulations at low mass scales are required to study how the growth of a baryonic component reshapes the dark matter halo and it is difficult to draw any strong conclusions from the observed discrepancy.

The model $\cos\eta$ distribution matches the weak preference for radial alignment from the simulations very nicely. This is in contrast to the results of dark-matter-only simulations \citep{Kiessling2015,Joachimi2015}, which produce satellite haloes that are significantly more radially aligned than observed in the data. The good agreement found here is possibly because we are using mock stellar samples generated from the most bound dark matter particles and it is known the degree of radial alignment decreases for the more bound particles \citep{PereiraBryan,Knebe2008b}. However, it should be acknowledged that even in the Aquarius catalogues there is a slightly larger number of the most radially-aligned haloes compared to the model although the results are consistent within the errors. There is also a slight overabundance of tangentially-aligned haloes in the simulations which may be produced by neglecting correlations between orbital phase and alignment \citep[cutting the sample on the absolute radial velocity produces a much flatter distribution in $\cos\eta$]{PereiraBryan} but doesn't seem to be produced by near oblate haloes with ill-defined major axis orientations. The match between the model and the Aquarius simulations leads us to conclude that the LG dSph population (in particular the ultrafaint population of M31 which drive the radial alignment signal) are aligned with their host haloes. Note also that the mass-weighted distribution is very similar to the non-mass-weighted distribution demonstrating that in the Aquarius simulations the degree of radial alignment is independent of mass \citep{Knebe2008a,Knebe2008b,PereiraBryan}.

\section{Velocity-dispersion against half-light radius scatter}\label{Sec::VelDisp}

\begin{figure*}
\includegraphics[width=\textwidth]{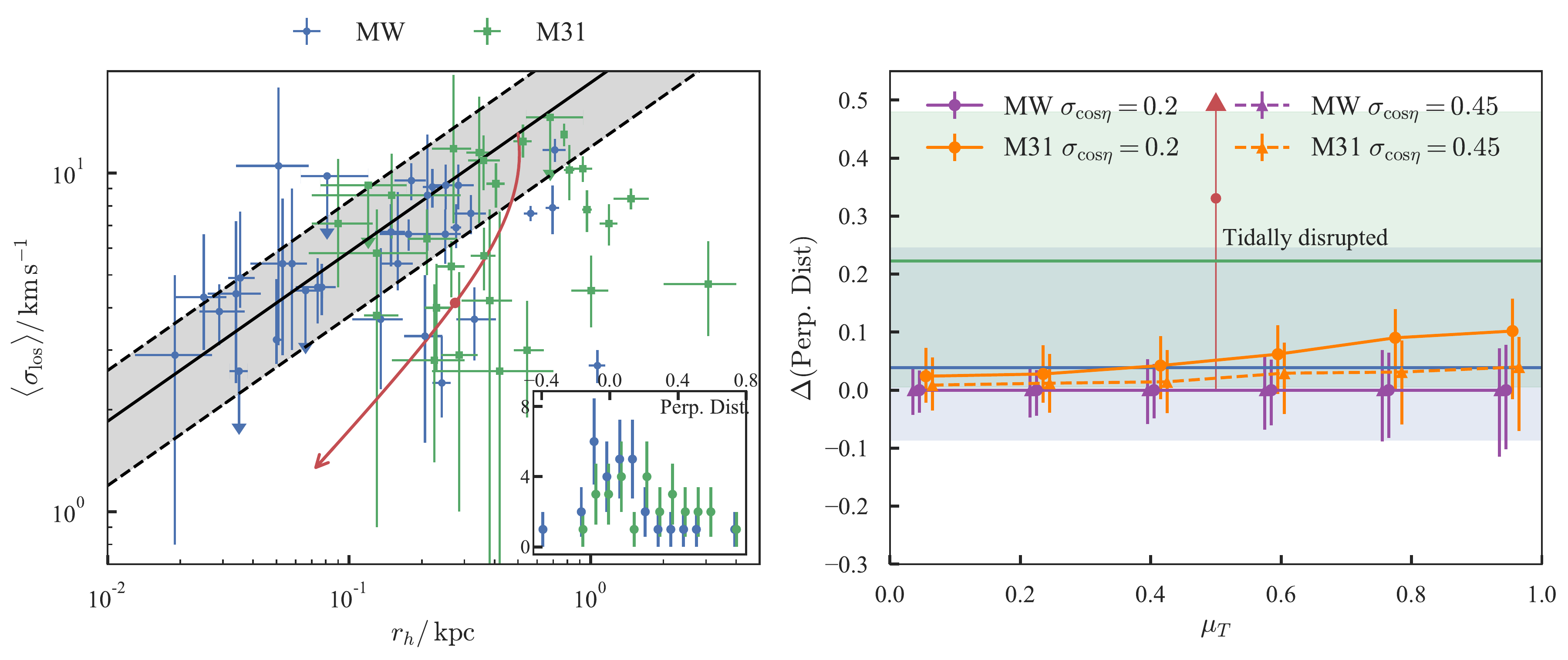}
\caption{Velocity dispersion against radius predictions: the \textbf{left panel} shows the line-of-sight velocity dispersion against projected major-axis length for Milky Way (MW) and Andromeda (M31) dwarf spheroidal galaxies (upper limits are shown as arrows). The black-line shows the relationship given by \protect\cite{Walker2010} with the grey band showing the $\pm 1\sigma$ interval. The red line shows the track from \protect\cite{Penarrubia2008} that a tidally-stripped dSph will follow whilst losing $99\percent$ of its stars with the red dot showing the point where $90\percent$ of the mass has been lost. The inset panel shows the distance (in log space) of the median velocity dispersion and half-light radius measurements with respect to the black line (with positive distance corresponding to points that fall beneath the line). The \textbf{right panel} shows model predictions of the discrepancy between the perpendicular distance distributions for M31 relative to the MW. We show two sets of models with varying mean triaxiality $\mu_T$ which differ in their degree of radial alignment ($\sigma_{\cos\eta}$). The two bands show the median (solid line) and $\pm1\sigma$ spread of the MW (blue) and M31 (green) data distributions.}
\label{Fig::DataPlot}
\end{figure*}

The dSphs seem to naturally fill in the low dispersion end of the scaling relation between radius and the circular velocity due to dark matter followed by data from spirals and low surface brightness galaxies \citep{Walker2010}. However, there appears to be significant scatter below this relationship with the dSphs of Andromeda falling further from the relation than those of the Milky Way. In the left panel of  Fig.~\ref{Fig::DataPlot}, we show the compilation of MW and M31 dSphs with luminosity-averaged line-of-sight velocity dispersions plotted against the length of their projected half-light major axis. In addition to the data given in the updated table from \cite{McConnachie2012}, we use the upper-bound velocity dispersion measurements for And XI from \cite{Collins2010} and for And XXIV from \cite{Collins2013} as well as the recent velocity dispersion measurements for And XXXI, And XXXII and And XXXIII from \cite{Martin2014}. We take velocity dispersion measurements for the recently-discovered MW dSphs from \cite{Torrealba2016a}, \cite{Caldwell2016}, \cite{Martin2016Draco}, \cite{Eridanus2017}, \cite{Walker2016}, \cite{Koposov2015}, \cite{Kirby2015}, \cite{Kim2016}, \cite{Kirby2013} and \cite{Martin2016triang}.
Also displayed is the velocity dispersion against radius relation followed by spiral and low surface brightness galaxies given by \cite{Walker2010}. The red line shows the path followed by a dwarf spheroidal undergoing tidal stripping from \cite{Penarrubia2008}. The inset panel shows the perpendicular logarithmic distance from the relation of \cite{Walker2010} (positive distance is defined to be the region beneath the line). Clearly, the Andromeda dSphs fall in general further below the line than the Milky Way dSphs. With our models of the dSph populations of M31 and MW, we now briefly investigate whether this effect could be due to projection effects.

If we assume that the dSph population of the MW and M31 follow similar shape-alignment distributions (as has been assumed throughout this paper), then one natural explanation for the difference in distance from the Walker relation is that we are observing the two populations from different perspectives and so preferentially observe the dSphs in certain directions. For the models from Section~\ref{Sec::Analysis}, we can generate and observe mock samples for M31 and MW. To do this, we require a sampling distribution for the positions of the dSphs. For simplicity we assume spherical symmetry. An exponential distribution in distance with a scale-length of $124\,\mathrm{kpc}$ and a truncation radius of $400\,\mathrm{kpc}$ gives a simple fit to the the MW dSph population that does not account for any incompleteness (for M31 we use a scaled version with a truncation radius of $350\mathrm{kpc}$ and $109\,\mathrm{kpc}$).

If we assume that the Walker relation traces the properties of galaxies averaged over the viewing angle, we hypothesise that any deviation from the relation is due to preferentially viewing along a particular direction. As discussed in \cite{SandersEvans2016}, under the assumption that the dark matter and stars are stratified on the same self-similar ellipsoids we can relate the ratio of the observed line-of-sight velocity dispersion to the spherically-averaged dispersion. Similarly, we can relate the observed major axis length of an ellipsoid to the equivalent spherical radius. For each of our samples, we calculate these ratios and hence the corresponding distance from the Walker relation. In the right panel of Fig.~\ref{Fig::DataPlot} we show the median M31 distance from the relation (with respect to the median MW distance) for a series of populations with $\mu_E=0.55$, $\sigma_E=0.15$ and $\sigma_T=0.05$ and varying $\mu_T$ for two different values of $\sigma_{\cos\eta}=0.2,0.45$ (approximately the smallest value consistent with the data and the maximum likelihood value). We see that for $\sigma_{\cos\eta}=0.45$ there is essentially no difference between the MW and M31 populations whilst for $\sigma_{\cos\eta}=0.2$ there is a weak trend with $\mu_T$ such that for the prolate figures ($\mu_T\sim1$) there is a difference in the median perpendicular distances in the same sense as the data. However, the magnitude of the difference is very small ($\sim0.1$) compared to the signal in the data. Note that more extreme differences can be created by using different shape and alignment distributions but such distributions are not consistent with the data. We therefore conclude that the scatter in the velocity dispersion against scale radius relation cannot be due to projection effects and that some other factor is at play. As noted in \cite{Walker2010}, perhaps the most convincing explanation is provided by tidal effects on the dSphs \citep{Penarrubia2008} which cause the dSphs to significantly stray from the Walker relation as shown in Fig.~\ref{Fig::DataPlot}. Tidal effects are naturally more significant in M31 due to its more massive disc -- \cite{McMillan2017} estimates the total stellar mass of the Milky Way as $M_*=(5.4\pm0.6)\times10^{10}M_\odot$ whilst \cite{Chemin2009} estimates the total stellar mass of M31 as $M_*=(9.5\pm1.7)\times10^{10}M_\odot$.

\section{Conclusions}\label{Sec::Conclusions}

The recent discoveries of low-luminosity dwarf galaxies in the Local Group (LG) have made it possible to begin analysing the collective properties of the dwarf spheroidal population. In particular, we are in the beneficial position of studying a dwarf spheroidal population both from approximately the centre of the galaxy in the Milky Way and from the outside of the galaxy for the case of Andromeda. We have analysed the intrinsic shape and alignment distributions of 33 Milky Way and 28 Andromeda dwarf spheroidals assuming the galaxies are all optically thin triaxial ellipsoids and compared the results to dark-matter-only simulations of a population of subhaloes. The conclusions of this study are as follows:

\begin{enumerate}
\item We fitted the observed ellipticities of the Milky Way (MW) and Andromeda (M31) dSphs assuming we do not observe the dSphs from any preferential direction. We found that the distribution of triaxialities is not well constrained whereas the intrinsic ellipticity distribution is well constrained as a Gaussian with mean $\mu_E=0.55\pm0.05$ and width $\sigma_E\approx0.15$. Examining the MW and M31 populations separately, we found that the MW dSphs ($\mu_E\approx0.6$) are intrinsically flatter than those of M31 ($\mu_E\approx0.5$). Splitting into classical ($M_V<-8.5\,\mathrm{mag}$) and ultrafaint populations, we find that the ultrafaints are flatter ($\mu_E\approx0.6$) than the classical dSphs ($\mu_E\approx0.5$). The classical populations of both M31 and the MW have very similar intrinsic ellipticity distributions whilst the discrepancy between the whole populations is driven by the ultrafaint dSphs.

\item We additionally used the distance moduli and position angles of the major axes of the dSphs to measure the distribution of the intrinsic alignments with respect to the galactocentric radial direction (of each dSph). We found that the Milky Way alone gives no constraint on the degree of radial alignment and that using the M31 dSphs (with and without the inclusion of the MW dSphs) we measure the spread of the misalignment about radial as approximately $45\deg$. This signal appears to be entirely driven by the ultrafaint M31 population. 

\item We compared these conclusions to the predictions from the tagged dark-matter-only Aquarius simulations provided by \cite{Lowing2015}. We found that the intrinsic ellipticities of Aquarius subhaloes are smaller than those of the LG dSphs with a difference in minor-to-major axis ratio of $\Delta(c/a)\approx0.4$. The degree of radial alignment, however, matches the results obtained from fitting the LG dSphs.

\item We also measured whether the dSph major axes preferentially lie in a plane. When using the M31 dwarfs, we find no indication of a preferred plane of alignment, despite claims in the literature of a Great Plane of Satellites (GPOS). Although \citet{Ibata2013} demonstrated the existence of an overdensity of satellites along a great circle through the centre of Andromeda, the satellites do not have their major axes preferentially aligned with this plane, as evident already from visual inspection of Fig.~\ref{Fig::M31}. This suggests the GPOS is not a long-lived structure but formed from recent group infall \citep{Bowden2014,Fernando2017} as there has not been the time for the major axes of the dSphs to become tidally-locked in the orbital plane. Discs as thin as observed in M31 cannot be long-lived in triaxial haloes~\citep{Bowden2013}, but, even assuming recent formation, explaining the observed thinness of the plane is challenging \citep{Metz2009,Angus2016}. A further possibility is that the GPOS is due to chance alignment \citep{Gillet2015} -- a hypothesis supported by evidence that the in-plane and out-of-plane satellites in the GPOS show no differences in their physical and morphological properties \cite[e.g.,][]{Collins2015}.

\item When using the Milky Way data (with or without the M31 data), we find that there {\it is} a preferred plane of alignment of the dSph major axes. It has a normal vector pointing towards $(\ell_n,b_n)=(127,5)\deg$. Using just the classical MW dSphs, we find $(\ell_n,b_n)=(137,5)\deg$ and using the ultrafaint MW dSphs we find $(\ell_n,b_n)=(115,2)\deg$. We note that this plane has a roughly similar orientation to the Vast Polar Orbital structure (VPOS) of \citet{Pawlowski2015}. The agreement is only crude, as the normal vectors lie $\sim30\deg$ apart on the sky. However, on analyzing the velocities of the classical dSphs, it appears that this plane too cannot be long-lived. Specifically, our posterior distributions permit us to compute the alignment between the normalized angular momentum vector of a dSph and the normal vector to our discovered plane. The resulting flattish distributions are not strongly peaked at vanishing misalignment, suggesting that the structure cannot survive over orbital timescales.

\item We closed by providing a perspective on the distribution of MW and M31 dSphs in the velocity dispersion against scale radius plane. The M31 dSphs have a much larger scatter about the observed relation given by \cite{Walker2010} than the MW dSphs. We demonstrate that if M31 and MW have the same dSph shape and alignment distributions, the projection effects can produce a small difference in the scatter about the \citeauthor{Walker2010} relation in the right sense but the amplitude is quite discrepant from that observed and we conclude the discrepancy must be due to the relative importance of tidal disruption in the two galaxies.
\end{enumerate}

\section*{Acknowledgments}
JLS is grateful to the Science and Technologies Facilities council for financial support. We thank Vasily Belokurov, Sergey Koposov and other menbers of the Cambridge Streams group for useful comments during the course of this project.

\bibliographystyle{mnras}
\bibliography{bibliography}

\appendix
\section{Models using ellipticity and position angle posterior samples}\label{Appendix}
For the models in this paper, we have used approximations to the distribution of ellipticity and position angle uncertainties which importantly neglect any correlation between these two quantities. For 21 M31 dSphs, \cite{Martin2016} has provided $250$ posterior samples of the ellipticity and position angle from the MCMC chains used to fit the photometry. For moderate ellipticities ($\epsilon\gtrsim0.2$) the correlations between ellipticity and position angle are small whereas for smaller ellipticities there is naturally a much larger scatter in the position angle estimates. We find that the simple approximations to the 1D distributions of ellipticity (described in Section~\ref{Sec::AsymmErrors}) match the distribution of the posterior samples.

As a check of the modelling procedure employed in the bulk of the paper, we also fit a model using the posterior samples for those dSphs analysed by \cite{Martin2016}. For this we employ a kernel density estimate for the posterior distributions of the ellipticity and position angle. For each dSph $m$, we write the $M$ ellipticity and position angle samples as $\{\bs{x}_i\}_m=\{(\epsilon_i,\mathrm{P.A.}_i)\}_m$ and we replace the distributions $\mathcal{N}'$ in equations~\eqref{Eqn::RandomModel} and~\eqref{Eqn::AlignModel} with the likelihood
\begin{equation}
\mathcal{L}_m=\frac{1}{M}\sum_i^M \frac{1}{\sqrt{2\pi\mathsf{\Sigma}_m}}\exp\Big[-\tfrac{1}{2}(\bs{x}_{im}-\bs{X}_m)^T\cdot\mathsf{\Sigma}^{-1}_m\cdot(\bs{x}_{im}-\bs{X}_m)\Big]
\end{equation}
where $\bs{X}_m$ are the parameters describing the `true' ellipticity and position angle of each dSph and $\mathsf{\Sigma}_m$ is the covariance matrix of the $\{x_i\}_m$ scaled by the bandwidth computed using Scott's rule.

\begin{figure}
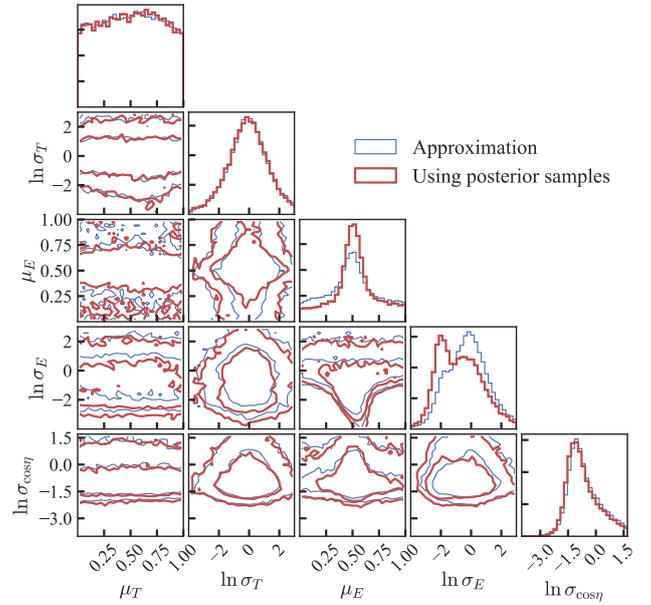

$$\includegraphics[width=\columnwidth]{{{figs/martin_samples_test_log}}}$$
\caption{Comparison of the radially-aligned inference using posterior samples from the error distributions from \protect\cite{Martin2016} (thick red) and using the approximation to the error distribution from Section~\ref{Sec::AsymmErrors} (thin blue).}
\label{Fig::SamplesModels}
\end{figure}

In Figure~\ref{Fig::SamplesModels}, we show the posteriors obtained by fitting this model compared to the model using the approximate distributions of Section~\ref{Sec::AsymmErrors} for the radially-aligned model case.

\label{lastpage}
\end{document}

%% file: data_latex.dat
Name&$\epsilon$&P.A.(deg)&$\mu$(mag)\\\hline Canes Venatici I&$0.39\pm0.03$&$70\pm4$&$21.69\pm0.10$\\Carina&$0.33\pm0.05$&$65\pm5$&$20.11\pm0.13$\\Draco&$0.31\pm0.02$&$89\pm2$&$19.40\pm0.17$\\Fornax&$0.30\pm0.01$&$41\pm1$&$20.84\pm0.18$\\Leo I&$0.21\pm0.03$&$79\pm3$&$22.02\pm0.13$\\Leo II&$0.13\pm0.05$&$12\pm10$&$21.84\pm0.13$\\Sculptor&$0.32\pm0.03$&$99\pm1$&$19.67\pm0.14$\\Sextans&$0.35\pm0.05$&$56\pm5$&$19.67\pm0.10$\\Ursa Minor&$0.56\pm0.05$&$53\pm5$&$19.40\pm0.10$\\\hline Aquarius II&$0.39\pm0.09$&$121\pm9$&$20.16\pm0.07$\\Bootes&$0.39\pm0.06$&$14\pm6$&$19.11\pm0.08$\\Bootes II&$<0.59$&$145\pm55$&$18.10\pm0.06$\\Canes Venatici II&$0.52\pm0.11$&$177\pm9$&$21.02\pm0.06$\\Coma Berenices&$0.38\pm0.14$&$115\pm10$&$18.20\pm0.20$\\Crater II&$<0.14$&-\footnote{For Crater II and Horologium I, the position angle is unconstrained.}&$20.35\pm0.02$\\Draco II&$<0.72$&$70\pm28$&$16.90\pm0.30$\\Eridanus II&$0.48\pm0.04$&$81\pm6$&$22.90\pm0.20$\\Grus I&$0.41^{+0.20}_{-0.28}$&$4\pm60$&$20.40\pm0.20$\\Hercules&$0.68\pm0.08$&$102\pm4$&$20.60\pm0.20$\\Horologium I&$<0.40$&-$^{a}$&$19.50\pm0.20$\\Hydra II&$<0.40$&$28^{+40}_{-35}$&$20.64\pm0.16$\\Leo IV&$0.49\pm0.11$&$121\pm9$&$20.94\pm0.09$\\Leo V&$0.50\pm0.15$&$96\pm13$&$21.25\pm0.12$\\Pegasus III&$<0.88$&$133\pm17$&$21.56\pm0.20$\\Pisces II&$0.40\pm0.10$&$77\pm12$&$21.30\pm0.20$\footnote{The distance modulus uncertainty for Pisces II is not provided by \cite{McConnachie2012} so we use a default value of $0.2$.}\\Reticulum II&$0.59^{+0.02}_{-0.03}$&$71\pm1$&$17.40\pm0.20$\\Segue I&$0.48\pm0.13$&$85\pm8$&$16.80\pm0.20$\\Segue II&$0.15\pm0.10$&$2\pm17$&$17.70\pm0.10$\\Triangulum II&$<0.54$&$56^{+16}_{-24}$&$17.40\pm0.10$\\Tucana II&$0.39^{+0.10}_{-0.20}$&$107\pm18$&$18.80\pm0.20$\\Ursa Major I&$0.80\pm0.04$&$71\pm3$&$19.93\pm0.10$\\Ursa Major II&$0.63\pm0.05$&$98\pm4$&$17.50\pm0.30$\\Willman I&$0.47\pm0.08$&$77\pm5$&$17.90\pm0.40$\\\hline And I&$0.28\pm0.03$&$30\pm4$&$24.36\pm0.07$\\And II&$0.16\pm0.02$&$31\pm5$&$24.07\pm0.06$\\And III&$0.59\pm0.04$&$140\pm3$&$24.37\pm0.07$\\And V&$0.26^{+0.09}_{-0.07}$&$54\pm10$&$24.44\pm0.08$\\And VI&$0.41\pm0.03$&$163\pm3$&$24.47\pm0.07$\\And VII&$0.13\pm0.04$&$94\pm8$&$24.41\pm0.10$\\And XIV&$<0.47$&$176\pm14$&$24.50^{+0.06}_{-0.56}$\\And XV&$0.24\pm0.10$&$38\pm15$&$23.98^{+0.26}_{-0.12}$\\And XVI&$0.29\pm0.08$&$98\pm9$&$23.39^{+0.19}_{-0.14}$\\And XVII&$0.50\pm0.10$&$110\pm9$&$24.31^{+0.11}_{-0.08}$\\And XIX&$0.58^{+0.05}_{-0.10}$&$34\pm5$&$24.57^{+0.08}_{-0.43}$\\And XXI&$0.36^{+0.10}_{-0.13}$&$139\pm13$&$24.59^{+0.06}_{-0.07}$\\And XXIII&$0.41^{+0.05}_{-0.06}$&$138\pm5$&$24.43\pm0.13$\\And XXV&$<0.33$&$164\pm30$&$24.55\pm0.12$\\And XXXI&$0.43\pm0.07$&$121\pm6$&$24.40\pm0.12$\\And XXXII&$0.50\pm0.09$&$90\pm7$&$24.45\pm0.14$\\And XXXIII&$0.43^{+0.15}_{-0.17}$&$174^{+15}_{-10}$&$24.44\pm0.18$\\\hline And IX&$<0.32$&$41\pm65$&$24.42\pm0.07$\\And X&$<0.71$&$30^{+20}_{-12}$&$24.13^{+0.08}_{-0.13}$\\And XI&$<0.68$&$54\pm30$&$24.33\pm0.05$\\And XII&$0.61^{+0.16}_{-0.48}$&$16^{+12}_{-36}$&$24.84^{+0.09}_{-0.34}$\\And XIII&$0.61^{+0.14}_{-0.20}$&$160^{+9}_{-12}$&$24.62\pm0.05$\\And XX&$<0.85$&$90^{+20}_{-44}$&$24.35^{+0.12}_{-0.16}$\\And XXII&$0.61^{+0.10}_{-0.14}$&$114\pm10$&$24.82^{+0.07}_{-0.36}$\\And XXIV&$<0.65$&$90\pm34$&$23.89\pm0.12$\\And XXVI&$<0.96$&$50\pm90$&$24.41\pm0.12$\\And XXIX&$0.35\pm0.06$&$51\pm8$&$24.32\pm0.22$\\And XXX&$0.43^{+0.10}_{-0.12}$&$110\pm9$&$24.17^{+0.10}_{-0.26}$\\\hline 

%% file: all_results.dat
\begin{tabular}{lllccccccc}\\&&&$\mu_T$&$\ln\,\sigma_T$&$\mu_E$&$\ln\,\sigma_E$&&&\\\hline\\Random&MW&All&$0.55^{+0.29}_{-0.36}$&$-0.10^{+1.27}_{-1.39}$&$0.61^{+0.20}_{-0.12}$&$-1.07^{+1.44}_{-0.84}$&&&\\&&&$(0.68)$&$(-0.04,-3.50)$&$(0.59)$&$(-1.73,-1.91)$&&&\\&&C&$0.52^{+0.32}_{-0.34}$&$-0.07^{+1.26}_{-1.31}$&$0.49^{+0.24}_{-0.16}$&$-1.03^{+1.70}_{-1.40}$&&&\\&&&$(0.63)$&$(-0.10,-3.67)$&$(0.47)$&$(-2.11,-2.46)$&&&\\&&UF&$0.56^{+0.30}_{-0.37}$&$0.01^{+1.24}_{-1.24}$&$0.65^{+0.22}_{-0.31}$&$-0.24^{+1.30}_{-1.12}$&&&\\&&&$(0.85)$&$(-0.02,1.47)$&$(0.65)$&$(-0.54,-1.92)$&&&\\&M31&All&$0.50^{+0.33}_{-0.33}$&$-0.04^{+1.23}_{-1.25}$&$0.52^{+0.10}_{-0.09}$&$-1.92^{+1.89}_{-0.84}$&&&\\&&&$(0.36)$&$(-0.20,-0.42)$&$(0.52)$&$(-2.16,-2.54)$&&&\\&&C&$0.51^{+0.32}_{-0.34}$&$-0.08^{+1.26}_{-1.35}$&$0.53^{+0.17}_{-0.14}$&$-1.40^{+1.85}_{-1.16}$&&&\\&&&$(0.58)$&$(-0.10,-3.75)$&$(0.52)$&$(-2.10,-2.58)$&&&\\&&UF&$0.50^{+0.34}_{-0.34}$&$-0.03^{+1.24}_{-1.24}$&$0.49^{+0.32}_{-0.31}$&$0.07^{+1.17}_{-1.08}$&&&\\&&&$(0.56)$&$(-0.02,-2.19)$&$(0.45)$&$(0.06,0.13)$&&&\\&Both&All&$0.54^{+0.30}_{-0.35}$&$-0.04^{+1.26}_{-1.27}$&$0.55^{+0.05}_{-0.04}$&$-1.94^{+0.36}_{-0.30}$&&&\\&&&$(0.67)$&$(-0.03,-2.73)$&$(0.55)$&$(-1.98,-2.04)$&&&\\&&C&$0.54^{+0.30}_{-0.36}$&$-0.11^{+1.27}_{-1.35}$&$0.50^{+0.05}_{-0.05}$&$-2.30^{+0.50}_{-0.53}$&&&\\&&&$(0.59)$&$(-0.11,-5.09)$&$(0.50)$&$(-2.26,-2.40)$&&&\\&&UF&$0.57^{+0.29}_{-0.38}$&$-0.07^{+1.26}_{-1.36}$&$0.62^{+0.23}_{-0.29}$&$-0.19^{+1.27}_{-1.13}$&&&\\&&&$(0.76)$&$(0.03,-2.81)$&$(0.62)$&$(-0.37,-1.76)$&&&\\\hline\\&&&$\mu_T$&$\ln\,\sigma_T$&$\mu_E$&$\ln\,\sigma_E$&$\ln\,\sigma_{\cos\eta}$&&\\\hline\\Aligned&MW&All&$0.51^{+0.32}_{-0.34}$&$-0.26^{+1.05}_{-1.41}$&$0.61^{+0.19}_{-0.11}$&$-1.18^{+1.29}_{-0.77}$&$-0.12^{+0.97}_{-1.18}$&&\\&&&$(0.58)$&$(-0.02,-3.67)$&$(0.59)$&$(-1.71,-1.94)$&$(-0.03,0.81)$&&\\&&C&$0.50^{+0.32}_{-0.34}$&$-0.24^{+1.04}_{-1.23}$&$0.50^{+0.22}_{-0.13}$&$-1.33^{+1.60}_{-1.23}$&$-0.12^{+0.97}_{-1.18}$&&\\&&&$(0.58)$&$(-0.25,-1.02)$&$(0.48)$&$(-2.12,-2.54)$&$(0.11,-1.81)$&&\\&&UF&$0.52^{+0.33}_{-0.36}$&$-0.17^{+1.01}_{-1.24}$&$0.66^{+0.22}_{-0.29}$&$-0.35^{+1.06}_{-1.02}$&$-0.18^{+1.00}_{-1.30}$&&\\&&&$(0.82)$&$(-0.02,-1.90)$&$(0.65)$&$(-0.66,-1.05)$&$(-0.04,-2.79)$&&\\&M31&All&$0.53^{+0.31}_{-0.35}$&$-0.22^{+1.02}_{-1.30}$&$0.52^{+0.11}_{-0.09}$&$-1.85^{+1.72}_{-0.83}$&$-0.24^{+0.94}_{-0.60}$&&\\&&&$(0.69)$&$(-0.03,-3.14)$&$(0.52)$&$(-2.12,-2.43)$&$(-0.65,-0.78)$&&\\&&C&$0.51^{+0.32}_{-0.33}$&$-0.20^{+1.01}_{-1.26}$&$0.54^{+0.18}_{-0.13}$&$-1.39^{+1.65}_{-1.15}$&$0.05^{+0.85}_{-0.69}$&&\\&&&$(0.56)$&$(-0.07,-1.67)$&$(0.53)$&$(-2.00,-2.47)$&$(-0.21,-0.50)$&&\\&&UF&$0.50^{+0.33}_{-0.34}$&$-0.17^{+0.99}_{-1.23}$&$0.49^{+0.31}_{-0.30}$&$-0.07^{+0.95}_{-1.08}$&$-0.91^{+1.26}_{-1.02}$&&\\&&&$(0.69)$&$(0.09,0.76)$&$(0.49)$&$(-0.03,1.28)$&$(-1.22,-1.91)$&&\\&Both&All&$0.52^{+0.31}_{-0.34}$&$-0.27^{+1.05}_{-1.35}$&$0.56^{+0.05}_{-0.04}$&$-1.93^{+0.37}_{-0.31}$&$-0.18^{+0.92}_{-0.63}$&&\\&&&$(0.69)$&$(-0.10,-3.26)$&$(0.56)$&$(-1.96,-2.04)$&$(-0.51,-0.72)$&&\\&&C&$0.55^{+0.30}_{-0.35}$&$-0.23^{+1.04}_{-1.28}$&$0.51^{+0.05}_{-0.05}$&$-2.31^{+0.51}_{-0.55}$&$0.02^{+0.86}_{-0.68}$&&\\&&&$(0.63)$&$(-0.10,-2.78)$&$(0.50)$&$(-2.27,-2.46)$&$(-0.36,-0.57)$&&\\&&UF&$0.48^{+0.34}_{-0.33}$&$-0.18^{+1.01}_{-1.24}$&$0.63^{+0.22}_{-0.28}$&$-0.30^{+1.04}_{-1.03}$&$-1.09^{+1.36}_{-1.03}$&&\\&&&$(0.07)$&$(-0.13,-4.08)$&$(0.62)$&$(-0.36,-1.05)$&$(-1.44,-2.11)$&&\\\hline\\&&&$\mu_T$&$\ln\,\sigma_T$&$\mu_E$&$\ln\,\sigma_E$&$\ell_n/\,\mathrm{deg}$&$\sin\,b_n$&$\ln\,\sigma_{\sin\alpha}$\\\hline\\Planar&MW&All&$0.51^{+0.31}_{-0.34}$&$-0.18^{+1.01}_{-1.24}$&$0.61^{+0.18}_{-0.10}$&$-1.28^{+1.28}_{-0.72}$&$127.46^{+10.06}_{-12.34}$&$0.07^{+0.20}_{-0.22}$&$-1.49^{+0.90}_{-0.87}$\\&&&$(0.60)$&$(-0.03,-2.95)$&$(0.59)$&$(-1.78,-1.95)$&$(128.87)$&$(0.09)$&$(-1.41,-1.95)$\\&&C&$0.52^{+0.31}_{-0.34}$&$-0.23^{+1.03}_{-1.30}$&$0.49^{+0.22}_{-0.13}$&$-1.41^{+1.68}_{-1.18}$&$136.83^{+22.81}_{-86.35}$&$0.09^{+0.38}_{-0.48}$&$-0.64^{+1.27}_{-1.25}$\\&&&$(0.62)$&$(-0.08,-3.23)$&$(0.47)$&$(-2.18,-2.52)$&$(146.44)$&$(0.12)$&$(-1.09,-1.97)$\\&&UF&$0.52^{+0.33}_{-0.36}$&$-0.18^{+1.01}_{-1.30}$&$0.66^{+0.21}_{-0.27}$&$-0.44^{+1.11}_{-1.08}$&$115.35^{+16.55}_{-30.78}$&$0.03^{+0.35}_{-0.39}$&$-0.96^{+1.42}_{-1.14}$\\&&&$(0.76)$&$(-0.08,-2.73)$&$(0.65)$&$(-0.64,-1.68)$&$(118.45)$&$(0.07)$&$(-1.30,-2.03)$\\&M31&All&$0.51^{+0.32}_{-0.34}$&$-0.22^{+1.03}_{-1.27}$&$0.52^{+0.09}_{-0.08}$&$-1.97^{+1.72}_{-0.83}$&$93.69^{+57.16}_{-61.76}$&$-0.18^{+0.77}_{-0.57}$&$0.14^{+0.83}_{-0.85}$\\&&&$(0.56)$&$(-0.07,-3.41)$&$(0.52)$&$(-2.19,-2.52)$&$(92.45)$&$(-0.76)$&$(0.08,1.10)$\\&&C&$0.51^{+0.32}_{-0.34}$&$-0.21^{+1.03}_{-1.28}$&$0.54^{+0.17}_{-0.12}$&$-1.47^{+1.69}_{-1.12}$&$90.27^{+61.67}_{-60.04}$&$-0.19^{+0.72}_{-0.51}$&$0.04^{+0.89}_{-0.97}$\\&&&$(0.64)$&$(-0.07,-3.57)$&$(0.53)$&$(-2.11,-2.50)$&$(85.16)$&$(-0.59)$&$(0.06,1.17)$\\&&UF&$0.49^{+0.34}_{-0.34}$&$-0.19^{+1.02}_{-1.26}$&$0.49^{+0.31}_{-0.31}$&$-0.08^{+0.98}_{-1.09}$&$98.90^{+52.05}_{-64.61}$&$0.00^{+0.73}_{-0.74}$&$-0.03^{+0.91}_{-1.05}$\\&&&$(0.58)$&$(0.04,-3.72)$&$(0.50)$&$(-0.11,1.09)$&$(136.13)$&$(-0.86)$&$(0.11,-3.52)$\\&Both&All&$0.50^{+0.32}_{-0.32}$&$-0.13^{+0.97}_{-1.16}$&$0.56^{+0.05}_{-0.04}$&$-1.95^{+0.38}_{-0.31}$&$127.49^{+11.73}_{-15.60}$&$0.05^{+0.21}_{-0.25}$&$-1.22^{+1.17}_{-0.84}$\\&&&$(0.42)$&$(0.02,-4.33)$&$(0.55)$&$(-1.99,-2.05)$&$(129.21)$&$(0.07)$&$(-1.32,-1.81)$\\&&C&$0.57^{+0.28}_{-0.36}$&$-0.24^{+1.04}_{-1.28}$&$0.50^{+0.05}_{-0.05}$&$-2.32^{+0.47}_{-0.52}$&$141.03^{+19.65}_{-88.51}$&$0.04^{+0.33}_{-0.45}$&$-0.56^{+1.21}_{-1.28}$\\&&&$(0.67)$&$(-0.05,-3.04)$&$(0.50)$&$(-2.26,-2.47)$&$(149.20)$&$(0.10)$&$(-0.79,-3.35)$\\&&UF&$0.50^{+0.33}_{-0.35}$&$-0.16^{+1.01}_{-1.26}$&$0.63^{+0.22}_{-0.26}$&$-0.39^{+1.08}_{-1.16}$&$115.54^{+17.18}_{-32.02}$&$0.04^{+0.37}_{-0.41}$&$-0.83^{+1.35}_{-1.11}$\\&&&$(0.68)$&$(0.04,1.28)$&$(0.62)$&$(-0.29,-1.96)$&$(118.43)$&$(0.06)$&$(-1.12,-1.88)$\\\hline\\\end{tabular}

%% file: main.bbl
\begin{thebibliography}{}
\makeatletter
\relax
\def\mn@urlcharsother{\let\do\@makeother \do\$\do\&\do\#\do\^\do\_\do\%\do\~}
\def\mn@doi{\begingroup\mn@urlcharsother \@ifnextchar [ {\mn@doi@}
  {\mn@doi@[]}}
\def\mn@doi@[#1]#2{\def\@tempa{#1}\ifx\@tempa\@empty \href
  {http://dx.doi.org/#2} {doi:#2}\else \href {http://dx.doi.org/#2} {#1}\fi
  \endgroup}
\def\mn@eprint#1#2{\mn@eprint@#1:#2::\@nil}
\def\mn@eprint@arXiv#1{\href {http://arxiv.org/abs/#1} {{\tt arXiv:#1}}}
\def\mn@eprint@dblp#1{\href {http://dblp.uni-trier.de/rec/bibtex/#1.xml}
  {dblp:#1}}
\def\mn@eprint@#1:#2:#3:#4\@nil{\def\@tempa {#1}\def\@tempb {#2}\def\@tempc
  {#3}\ifx \@tempc \@empty \let \@tempc \@tempb \let \@tempb \@tempa \fi \ifx
  \@tempb \@empty \def\@tempb {arXiv}\fi \@ifundefined
  {mn@eprint@\@tempb}{\@tempb:\@tempc}{\expandafter \expandafter \csname
  mn@eprint@\@tempb\endcsname \expandafter{\@tempc}}}

\bibitem[\protect\citeauthoryear{{Allgood}, {Flores}, {Primack}, {Kravtsov},
  {Wechsler}, {Faltenbacher}  \& {Bullock}}{{Allgood}
  et~al.}{2006}]{Allgood2006}
{Allgood} B.,  {Flores} R.~A.,  {Primack} J.~R.,  {Kravtsov} A.~V.,  {Wechsler}
  R.~H.,  {Faltenbacher} A.,   {Bullock} J.~S.,  2006, \mn@doi [\mnras]
  {10.1111/j.1365-2966.2006.10094.x}, \href
  {http://adsabs.harvard.edu/abs/2006MNRAS.367.1781A} {367, 1781}

\bibitem[\protect\citeauthoryear{{Angus}, {Coppin}, {Gentile}  \&
  {Diaferio}}{{Angus} et~al.}{2016}]{Angus2016}
{Angus} G.~W.,  {Coppin} P.,  {Gentile} G.,   {Diaferio} A.,  2016, \mn@doi
  [\mnras] {10.1093/mnras/stw1822}, \href
  {http://adsabs.harvard.edu/abs/2016MNRAS.462.3221A} {462, 3221}

\bibitem[\protect\citeauthoryear{{Bailin} \& {Steinmetz}}{{Bailin} \&
  {Steinmetz}}{2005}]{BailinSteinmetz2005}
{Bailin} J.,  {Steinmetz} M.,  2005, \mn@doi [\apj] {10.1086/430397}, \href
  {http://adsabs.harvard.edu/abs/2005ApJ...627..647B} {627, 647}

\bibitem[\protect\citeauthoryear{{Bailin}, {Bell}, {Valluri}, {Stinson},
  {Debattista}, {Couchman}  \& {Wadsley}}{{Bailin} et~al.}{2014}]{Bailin2014}
{Bailin} J.,  {Bell} E.~F.,  {Valluri} M.,  {Stinson} G.~S.,  {Debattista}
  V.~P.,  {Couchman} H.~M.~P.,   {Wadsley} J.,  2014, \mn@doi [\apj]
  {10.1088/0004-637X/783/2/95}, \href
  {http://adsabs.harvard.edu/abs/2014ApJ...783...95B} {783, 95}

\bibitem[\protect\citeauthoryear{{Barber}, {Starkenburg}, {Navarro}  \&
  {McConnachie}}{{Barber} et~al.}{2015}]{Barber2015}
{Barber} C.,  {Starkenburg} E.,  {Navarro} J.~F.,   {McConnachie} A.~W.,  2015,
  \mn@doi [\mnras] {10.1093/mnras/stu2494}, \href
  {http://adsabs.harvard.edu/abs/2015MNRAS.447.1112B} {447, 1112}

\bibitem[\protect\citeauthoryear{{Barlow}}{{Barlow}}{2004}]{Barlow2004}
{Barlow} R.,  2004, preprint, \href
  {http://adsabs.harvard.edu/abs/2004physics...6120B} {} (\mn@eprint {arXiv}
  {physics/0406120})

\bibitem[\protect\citeauthoryear{{Binggeli}}{{Binggeli}}{1980}]{Binggeli1980}
{Binggeli} B.,  1980, \aap, \href
  {http://adsabs.harvard.edu/abs/1980A%26A....82..289B} {82, 289}

\bibitem[\protect\citeauthoryear{{Binney}}{{Binney}}{1985}]{Binney1985}
{Binney} J.,  1985, \mn@doi [\mnras] {10.1093/mnras/212.4.767}, \href
  {http://adsabs.harvard.edu/abs/1985MNRAS.212..767B} {212, 767}

\bibitem[\protect\citeauthoryear{{Bowden}, {Evans}  \& {Belokurov}}{{Bowden}
  et~al.}{2013}]{Bowden2013}
{Bowden} A.,  {Evans} N.~W.,   {Belokurov} V.,  2013, \mn@doi [\mnras]
  {10.1093/mnras/stt1253}, \href
  {http://adsabs.harvard.edu/abs/2013MNRAS.435..928B} {435, 928}

\bibitem[\protect\citeauthoryear{{Bowden}, {Evans}  \& {Belokurov}}{{Bowden}
  et~al.}{2014}]{Bowden2014}
{Bowden} A.,  {Evans} N.~W.,   {Belokurov} V.,  2014, \mn@doi [\apjl]
  {10.1088/2041-8205/793/2/L42}, \href
  {http://adsabs.harvard.edu/abs/2014ApJ...793L..42B} {793, L42}

\bibitem[\protect\citeauthoryear{{Caldwell} et~al.,}{{Caldwell}
  et~al.}{2017}]{Caldwell2016}
{Caldwell} N.,  et~al., 2017, \mn@doi [\apj] {10.3847/1538-4357/aa688e}, \href
  {/abs/2017ApJ...839...20C} {839, 20}

\bibitem[\protect\citeauthoryear{{Chemin}, {Carignan}  \& {Foster}}{{Chemin}
  et~al.}{2009}]{Chemin2009}
{Chemin} L.,  {Carignan} C.,   {Foster} T.,  2009, \mn@doi [\apj]
  {10.1088/0004-637X/705/2/1395}, \href
  {http://adsabs.harvard.edu/abs/2009ApJ...705.1395C} {705, 1395}

\bibitem[\protect\citeauthoryear{{Collins} et~al.,}{{Collins}
  et~al.}{2010}]{Collins2010}
{Collins} M.~L.~M.,  et~al., 2010, \mn@doi [\mnras]
  {10.1111/j.1365-2966.2010.17069.x}, \href
  {http://adsabs.harvard.edu/abs/2010MNRAS.407.2411C} {407, 2411}

\bibitem[\protect\citeauthoryear{{Collins} et~al.,}{{Collins}
  et~al.}{2013}]{Collins2013}
{Collins} M.~L.~M.,  et~al., 2013, \mn@doi [\apj]
  {10.1088/0004-637X/768/2/172}, \href
  {http://adsabs.harvard.edu/abs/2013ApJ...768..172C} {768, 172}

\bibitem[\protect\citeauthoryear{{Collins} et~al.,}{{Collins}
  et~al.}{2015}]{Collins2015}
{Collins} M.~L.~M.,  et~al., 2015, \mn@doi [\apjl]
  {10.1088/2041-8205/799/1/L13}, \href
  {http://adsabs.harvard.edu/abs/2015ApJ...799L..13C} {799, L13}

\bibitem[\protect\citeauthoryear{{Contopoulos}}{{Contopoulos}}{1956}]{Contopoulos1956}
{Contopoulos} G.,  1956, \zap, \href
  {http://adsabs.harvard.edu/abs/1956ZA.....39..126C} {39, 126}

\bibitem[\protect\citeauthoryear{{Cooper} et~al.,}{{Cooper}
  et~al.}{2010}]{Cooper2010}
{Cooper} A.~P.,  et~al., 2010, \mn@doi [\mnras]
  {10.1111/j.1365-2966.2010.16740.x}, \href
  {http://adsabs.harvard.edu/abs/2010MNRAS.406..744C} {406, 744}

\bibitem[\protect\citeauthoryear{{Cooper}, {Cole}, {Frenk}, {Le Bret}  \&
  {Pontzen}}{{Cooper} et~al.}{2017}]{Cooper2016}
{Cooper} A.~P.,  {Cole} S.,  {Frenk} C.~S.,  {Le Bret} T.,   {Pontzen} A.,
  2017, \mn@doi [\mnras] {10.1093/mnras/stx955}, \href
  {http://adsabs.harvard.edu/abs/2017MNRAS.469.1691C} {469, 1691}

\bibitem[\protect\citeauthoryear{{D'Onghia}, {Springel}, {Hernquist}  \&
  {Keres}}{{D'Onghia} et~al.}{2010}]{DOnghia2010}
{D'Onghia} E.,  {Springel} V.,  {Hernquist} L.,   {Keres} D.,  2010, \mn@doi
  [\apj] {10.1088/0004-637X/709/2/1138}, \href
  {http://adsabs.harvard.edu/abs/2010ApJ...709.1138D} {709, 1138}

\bibitem[\protect\citeauthoryear{{Debattista}, {Moore}, {Quinn}, {Kazantzidis},
  {Maas}, {Mayer}, {Read}  \& {Stadel}}{{Debattista}
  et~al.}{2008}]{Debattista2008}
{Debattista} V.~P.,  {Moore} B.,  {Quinn} T.,  {Kazantzidis} S.,  {Maas} R.,
  {Mayer} L.,  {Read} J.,   {Stadel} J.,  2008, \mn@doi [\apj]
  {10.1086/587977}, \href {http://adsabs.harvard.edu/abs/2008ApJ...681.1076D}
  {681, 1076}

\bibitem[\protect\citeauthoryear{{Diaz}, {Koposov}, {Irwin}, {Belokurov}  \&
  {Evans}}{{Diaz} et~al.}{2014}]{Diaz2014}
{Diaz} J.~D.,  {Koposov} S.~E.,  {Irwin} M.,  {Belokurov} V.,   {Evans} N.~W.,
  2014, \mn@doi [\mnras] {10.1093/mnras/stu1210}, \href
  {http://adsabs.harvard.edu/abs/2014MNRAS.443.1688D} {443, 1688}

\bibitem[\protect\citeauthoryear{{Diemand}, {Kuhlen}  \& {Madau}}{{Diemand}
  et~al.}{2007}]{Diemand2007}
{Diemand} J.,  {Kuhlen} M.,   {Madau} P.,  2007, \mn@doi [\apj]
  {10.1086/510736}, \href {http://adsabs.harvard.edu/abs/2007ApJ...657..262D}
  {657, 262}

\bibitem[\protect\citeauthoryear{{Evans}, {Carollo}  \& {de Zeeuw}}{{Evans}
  et~al.}{2000}]{Evans2000}
{Evans} N.~W.,  {Carollo} C.~M.,   {de Zeeuw} P.~T.,  2000, \mn@doi [\mnras]
  {10.1046/j.1365-8711.2000.03787.x}, \href
  {http://adsabs.harvard.edu/abs/2000MNRAS.318.1131E} {318, 1131}

\bibitem[\protect\citeauthoryear{{Faltenbacher}, {Jing}, {Li}, {Mao}, {Mo},
  {Pasquali}  \& {van den Bosch}}{{Faltenbacher}
  et~al.}{2008}]{Faltenbacher2008}
{Faltenbacher} A.,  {Jing} Y.~P.,  {Li} C.,  {Mao} S.,  {Mo} H.~J.,  {Pasquali}
  A.,   {van den Bosch} F.~C.,  2008, \mn@doi [\apj] {10.1086/525243}, \href
  {http://adsabs.harvard.edu/abs/2008ApJ...675..146F} {675, 146}

\bibitem[\protect\citeauthoryear{{Fernando}, {Arias}, {Guglielmo}, {Lewis},
  {Ibata}  \& {Power}}{{Fernando} et~al.}{2017}]{Fernando2017}
{Fernando} N.,  {Arias} V.,  {Guglielmo} M.,  {Lewis} G.~F.,  {Ibata} R.~A.,
  {Power} C.,  2017, \mn@doi [\mnras] {10.1093/mnras/stw2694}, \href
  {http://adsabs.harvard.edu/abs/2017MNRAS.465..641F} {465, 641}

\bibitem[\protect\citeauthoryear{{Font} et~al.,}{{Font}
  et~al.}{2011}]{Font2011}
{Font} A.~S.,  et~al., 2011, \mn@doi [\mnras]
  {10.1111/j.1365-2966.2011.19339.x}, \href
  {http://adsabs.harvard.edu/abs/2011MNRAS.417.1260F} {417, 1260}

\bibitem[\protect\citeauthoryear{{Franx}, {Illingworth}  \& {de Zeeuw}}{{Franx}
  et~al.}{1991}]{Franx1991}
{Franx} M.,  {Illingworth} G.,   {de Zeeuw} T.,  1991, \mn@doi [\apj]
  {10.1086/170769}, \href {http://adsabs.harvard.edu/abs/1991ApJ...383..112F}
  {383, 112}

\bibitem[\protect\citeauthoryear{Gelman et~al.}{Gelman
  et~al.}{2006}]{Gelman2006}
Gelman A.,  et~al., 2006, Bayesian analysis, 1, 515

\bibitem[\protect\citeauthoryear{{Gerhard}}{{Gerhard}}{1983}]{Gerhard1983}
{Gerhard} O.~E.,  1983, \mn@doi [\mnras] {10.1093/mnras/202.4.1159}, \href
  {http://adsabs.harvard.edu/abs/1983MNRAS.202.1159G} {202, 1159}

\bibitem[\protect\citeauthoryear{{Gillet}, {Ocvirk}, {Aubert}, {Knebe},
  {Libeskind}, {Yepes}, {Gottl{\"o}ber}  \& {Hoffman}}{{Gillet}
  et~al.}{2015}]{Gillet2015}
{Gillet} N.,  {Ocvirk} P.,  {Aubert} D.,  {Knebe} A.,  {Libeskind} N.,  {Yepes}
  G.,  {Gottl{\"o}ber} S.,   {Hoffman} Y.,  2015, \mn@doi [\apj]
  {10.1088/0004-637X/800/1/34}, \href
  {http://adsabs.harvard.edu/abs/2015ApJ...800...34G} {800, 34}

\bibitem[\protect\citeauthoryear{Hoffman \& Gelman}{Hoffman \&
  Gelman}{2014}]{NUTSsampler}
Hoffman M.~D.,  Gelman A.,  2014, Journal of Machine Learning Research, 15,
  1593

\bibitem[\protect\citeauthoryear{{Ibata} et~al.,}{{Ibata}
  et~al.}{2013}]{Ibata2013}
{Ibata} R.~A.,  et~al., 2013, \mn@doi [\nat] {10.1038/nature11717}, \href
  {http://adsabs.harvard.edu/abs/2013Natur.493...62I} {493, 62}

\bibitem[\protect\citeauthoryear{{Jing} \& {Suto}}{{Jing} \&
  {Suto}}{2002}]{JingSuto2002}
{Jing} Y.~P.,  {Suto} Y.,  2002, \mn@doi [\apj] {10.1086/341065}, \href
  {http://adsabs.harvard.edu/abs/2002ApJ...574..538J} {574, 538}

\bibitem[\protect\citeauthoryear{{Joachimi} et~al.,}{{Joachimi}
  et~al.}{2015}]{Joachimi2015}
{Joachimi} B.,  et~al., 2015, \mn@doi [\ssr] {10.1007/s11214-015-0177-4}, \href
  {http://adsabs.harvard.edu/abs/2015SSRv..193....1J} {193, 1}

\bibitem[\protect\citeauthoryear{{Kiessling} et~al.,}{{Kiessling}
  et~al.}{2015}]{Kiessling2015}
{Kiessling} A.,  et~al., 2015, \mn@doi [\ssr] {10.1007/s11214-015-0203-6},
  \href {http://adsabs.harvard.edu/abs/2015SSRv..193...67K} {193, 67}

\bibitem[\protect\citeauthoryear{{Kim} et~al.,}{{Kim} et~al.}{2016}]{Kim2016}
{Kim} D.,  et~al., 2016, \mn@doi [\apj] {10.3847/0004-637X/833/1/16}, \href
  {http://adsabs.harvard.edu/abs/2016ApJ...833...16K} {833, 16}

\bibitem[\protect\citeauthoryear{{Kirby}, {Boylan-Kolchin}, {Cohen}, {Geha},
  {Bullock}  \& {Kaplinghat}}{{Kirby} et~al.}{2013}]{Kirby2013}
{Kirby} E.~N.,  {Boylan-Kolchin} M.,  {Cohen} J.~G.,  {Geha} M.,  {Bullock}
  J.~S.,   {Kaplinghat} M.,  2013, \mn@doi [\apj] {10.1088/0004-637X/770/1/16},
  \href {http://adsabs.harvard.edu/abs/2013ApJ...770...16K} {770, 16}

\bibitem[\protect\citeauthoryear{{Kirby}, {Simon}  \& {Cohen}}{{Kirby}
  et~al.}{2015}]{Kirby2015}
{Kirby} E.~N.,  {Simon} J.~D.,   {Cohen} J.~G.,  2015, \mn@doi [\apj]
  {10.1088/0004-637X/810/1/56}, \href
  {http://adsabs.harvard.edu/abs/2015ApJ...810...56K} {810, 56}

\bibitem[\protect\citeauthoryear{{Kleyna}, {Wilkinson}, {Evans}  \&
  {Gilmore}}{{Kleyna} et~al.}{2001}]{Kleyna2001}
{Kleyna} J.~T.,  {Wilkinson} M.~I.,  {Evans} N.~W.,   {Gilmore} G.,  2001,
  \mn@doi [\apjl] {10.1086/338603}, \href
  {http://adsabs.harvard.edu/abs/2001ApJ...563L.115K} {563, L115}

\bibitem[\protect\citeauthoryear{{Knebe}, {Draganova}, {Power}, {Yepes},
  {Hoffman}, {Gottl{\"o}ber}  \& {Gibson}}{{Knebe} et~al.}{2008a}]{Knebe2008a}
{Knebe} A.,  {Draganova} N.,  {Power} C.,  {Yepes} G.,  {Hoffman} Y.,
  {Gottl{\"o}ber} S.,   {Gibson} B.~K.,  2008a, \mn@doi [\mnras]
  {10.1111/j.1745-3933.2008.00459.x}, \href
  {http://adsabs.harvard.edu/abs/2008MNRAS.386L..52K} {386, L52}

\bibitem[\protect\citeauthoryear{{Knebe}, {Yahagi}, {Kase}, {Lewis}  \&
  {Gibson}}{{Knebe} et~al.}{2008b}]{Knebe2008b}
{Knebe} A.,  {Yahagi} H.,  {Kase} H.,  {Lewis} G.,   {Gibson} B.~K.,  2008b,
  \mn@doi [\mnras] {10.1111/j.1745-3933.2008.00495.x}, \href
  {http://adsabs.harvard.edu/abs/2008MNRAS.388L..34K} {388, L34}

\bibitem[\protect\citeauthoryear{{Knebe}, {Libeskind}, {Knollmann}, {Yepes},
  {Gottl{\"o}ber}  \& {Hoffman}}{{Knebe} et~al.}{2010}]{Knebe2010}
{Knebe} A.,  {Libeskind} N.~I.,  {Knollmann} S.~R.,  {Yepes} G.,
  {Gottl{\"o}ber} S.,   {Hoffman} Y.,  2010, \mn@doi [\mnras]
  {10.1111/j.1365-2966.2010.16514.x}, \href
  {http://adsabs.harvard.edu/abs/2010MNRAS.405.1119K} {405, 1119}

\bibitem[\protect\citeauthoryear{{Koposov} et~al.,}{{Koposov}
  et~al.}{2015}]{Koposov2015}
{Koposov} S.~E.,  et~al., 2015, \mn@doi [\apj] {10.1088/0004-637X/811/1/62},
  \href {http://adsabs.harvard.edu/abs/2015ApJ...811...62K} {811, 62}

\bibitem[\protect\citeauthoryear{{Kuhlen}, {Diemand}  \& {Madau}}{{Kuhlen}
  et~al.}{2007}]{Kuhlen2007}
{Kuhlen} M.,  {Diemand} J.,   {Madau} P.,  2007, \mn@doi [\apj]
  {10.1086/522878}, \href {http://adsabs.harvard.edu/abs/2007ApJ...671.1135K}
  {671, 1135}

\bibitem[\protect\citeauthoryear{{Le Bret}, {Pontzen}, {Cooper}, {Frenk},
  {Zolotov}, {Brooks}, {Governato}  \& {Parry}}{{Le Bret}
  et~al.}{2015}]{LeBret2015}
{Le Bret} T.,  {Pontzen} A.,  {Cooper} A.~P.,  {Frenk} C.,  {Zolotov} A.,
  {Brooks} A.~M.,  {Governato} F.,   {Parry} O.~H.,  2015, preprint, \href
  {http://adsabs.harvard.edu/abs/2015arXiv150206371L} {} (\mn@eprint {arXiv}
  {1502.06371})

\bibitem[\protect\citeauthoryear{{Li} et~al.,}{{Li}
  et~al.}{2017}]{Eridanus2017}
{Li} T.~S.,  et~al., 2017, \mn@doi [\apj] {10.3847/1538-4357/aa6113}, \href
  {http://adsabs.harvard.edu/abs/2017ApJ...838....8L} {838, 8}

\bibitem[\protect\citeauthoryear{{Lowing}, {Wang}, {Cooper}, {Kennedy},
  {Helly}, {Cole}  \& {Frenk}}{{Lowing} et~al.}{2015}]{Lowing2015}
{Lowing} B.,  {Wang} W.,  {Cooper} A.,  {Kennedy} R.,  {Helly} J.,  {Cole} S.,
   {Frenk} C.,  2015, \mn@doi [\mnras] {10.1093/mnras/stu2257}, \href
  {http://adsabs.harvard.edu/abs/2015MNRAS.446.2274L} {446, 2274}

\bibitem[\protect\citeauthoryear{{Lynden-Bell}}{{Lynden-Bell}}{1976}]{LyndenBell1976}
{Lynden-Bell} D.,  1976, \mn@doi [\mnras] {10.1093/mnras/174.3.695}, \href
  {http://adsabs.harvard.edu/abs/1976MNRAS.174..695L} {174, 695}

\bibitem[\protect\citeauthoryear{{Lynden-Bell}}{{Lynden-Bell}}{1994}]{LyndenBell1994}
{Lynden-Bell} D.,  1994, in {Meylan} G.,  {Prugniel} P.,  eds,  European
  Southern Observatory Conference and Workshop Proceedings Vol. 49, European
  Southern Observatory Conference and Workshop Proceedings. p.~589

\bibitem[\protect\citeauthoryear{{Lynden-Bell} \& {Lynden-Bell}}{{Lynden-Bell}
  \& {Lynden-Bell}}{1995}]{LyndenBell1995}
{Lynden-Bell} D.,  {Lynden-Bell} R.~M.,  1995, \mn@doi [\mnras]
  {10.1093/mnras/275.2.429}, \href
  {http://adsabs.harvard.edu/abs/1995MNRAS.275..429L} {275, 429}

\bibitem[\protect\citeauthoryear{{Martin}, {de Jong}  \& {Rix}}{{Martin}
  et~al.}{2008}]{Martin2008}
{Martin} N.~F.,  {de Jong} J.~T.~A.,   {Rix} H.-W.,  2008, \mn@doi [\apj]
  {10.1086/590336}, \href {http://adsabs.harvard.edu/abs/2008ApJ...684.1075M}
  {684, 1075}

\bibitem[\protect\citeauthoryear{{Martin} et~al.,}{{Martin}
  et~al.}{2014}]{Martin2014}
{Martin} N.~F.,  et~al., 2014, \mn@doi [\apjl] {10.1088/2041-8205/793/1/L14},
  \href {http://adsabs.harvard.edu/abs/2014ApJ...793L..14M} {793, L14}

\bibitem[\protect\citeauthoryear{{Martin} et~al.,}{{Martin}
  et~al.}{2016a}]{Martin2016Draco}
{Martin} N.~F.,  et~al., 2016a, \mn@doi [\mnras] {10.1093/mnrasl/slw013}, \href
  {http://adsabs.harvard.edu/abs/2016MNRAS.458L..59M} {458, L59}

\bibitem[\protect\citeauthoryear{{Martin} et~al.,}{{Martin}
  et~al.}{2016b}]{Martin2016triang}
{Martin} N.~F.,  et~al., 2016b, \mn@doi [\apj] {10.3847/0004-637X/818/1/40},
  \href {http://adsabs.harvard.edu/abs/2016ApJ...818...40M} {818, 40}

\bibitem[\protect\citeauthoryear{{Martin} et~al.,}{{Martin}
  et~al.}{2016c}]{Martin2016}
{Martin} N.~F.,  et~al., 2016c, \mn@doi [\apj] {10.3847/1538-4357/833/2/167},
  \href {http://adsabs.harvard.edu/abs/2016ApJ...833..167M} {833, 167}

\bibitem[\protect\citeauthoryear{{Mateo}}{{Mateo}}{1998}]{Mateo1998}
{Mateo} M.~L.,  1998, \mn@doi [\araa] {10.1146/annurev.astro.36.1.435}, \href
  {http://adsabs.harvard.edu/abs/1998ARA%26A..36..435M} {36, 435}

\bibitem[\protect\citeauthoryear{{McConnachie}}{{McConnachie}}{2012}]{McConnachie2012}
{McConnachie} A.~W.,  2012, \mn@doi [\aj] {10.1088/0004-6256/144/1/4}, \href
  {http://adsabs.harvard.edu/abs/2012AJ....144....4M} {144, 4}

\bibitem[\protect\citeauthoryear{{McMillan}}{{McMillan}}{2017}]{McMillan2017}
{McMillan} P.~J.,  2017, \mn@doi [\mnras] {10.1093/mnras/stw2759}, \href
  {http://adsabs.harvard.edu/abs/2017MNRAS.465...76M} {465, 76}

\bibitem[\protect\citeauthoryear{{Metz}, {Kroupa}, {Theis}, {Hensler}  \&
  {Jerjen}}{{Metz} et~al.}{2009}]{Metz2009}
{Metz} M.,  {Kroupa} P.,  {Theis} C.,  {Hensler} G.,   {Jerjen} H.,  2009,
  \mn@doi [\apj] {10.1088/0004-637X/697/1/269}, \href
  {http://adsabs.harvard.edu/abs/2009ApJ...697..269M} {697, 269}

\bibitem[\protect\citeauthoryear{{Pawlowski} \& {Kroupa}}{{Pawlowski} \&
  {Kroupa}}{2013}]{Pawlowski2013}
{Pawlowski} M.~S.,  {Kroupa} P.,  2013, \mn@doi [\mnras]
  {10.1093/mnras/stt1429}, \href
  {http://adsabs.harvard.edu/abs/2013MNRAS.435.2116P} {435, 2116}

\bibitem[\protect\citeauthoryear{{Pawlowski}, {McGaugh}  \&
  {Jerjen}}{{Pawlowski} et~al.}{2015}]{Pawlowski2015}
{Pawlowski} M.~S.,  {McGaugh} S.~S.,   {Jerjen} H.,  2015, \mn@doi [\mnras]
  {10.1093/mnras/stv1588}, \href
  {http://adsabs.harvard.edu/abs/2015MNRAS.453.1047P} {453, 1047}

\bibitem[\protect\citeauthoryear{{Pe{\~n}arrubia}, {Navarro}  \&
  {McConnachie}}{{Pe{\~n}arrubia} et~al.}{2008}]{Penarrubia2008}
{Pe{\~n}arrubia} J.,  {Navarro} J.~F.,   {McConnachie} A.~W.,  2008, \mn@doi
  [\apj] {10.1086/523686}, \href
  {http://adsabs.harvard.edu/abs/2008ApJ...673..226P} {673, 226}

\bibitem[\protect\citeauthoryear{{Pereira} \& {Bryan}}{{Pereira} \&
  {Bryan}}{2010}]{PereiraBryan2010}
{Pereira} M.~J.,  {Bryan} G.~L.,  2010, \mn@doi [\apj]
  {10.1088/0004-637X/721/2/939}, \href
  {http://adsabs.harvard.edu/abs/2010ApJ...721..939P} {721, 939}

\bibitem[\protect\citeauthoryear{{Pereira}, {Bryan}  \& {Gill}}{{Pereira}
  et~al.}{2008}]{PereiraBryan}
{Pereira} M.~J.,  {Bryan} G.~L.,   {Gill} S.~P.~D.,  2008, \mn@doi [\apj]
  {10.1086/523830}, \href {http://adsabs.harvard.edu/abs/2008ApJ...672..825P}
  {672, 825}

\bibitem[\protect\citeauthoryear{{Salomon}, {Ibata}, {Martin}  \&
  {Famaey}}{{Salomon} et~al.}{2015}]{Salomon2015}
{Salomon} J.-B.,  {Ibata} R.~A.,  {Martin} N.~F.,   {Famaey} B.,  2015, \mn@doi
  [\mnras] {10.1093/mnras/stv741}, \href
  {http://adsabs.harvard.edu/abs/2015MNRAS.450.1409S} {450, 1409}

\bibitem[\protect\citeauthoryear{{S{\'a}nchez-Janssen}
  et~al.,}{{S{\'a}nchez-Janssen} et~al.}{2016}]{SanchezJanssen}
{S{\'a}nchez-Janssen} R.,  et~al., 2016, \mn@doi [\apj]
  {10.3847/0004-637X/820/1/69}, \href
  {http://adsabs.harvard.edu/abs/2016ApJ...820...69S} {820, 69}

\bibitem[\protect\citeauthoryear{{Sanders} \& {Evans}}{{Sanders} \&
  {Evans}}{2016}]{SandersEvans2016}
{Sanders} J.~L.,  {Evans} N.~W.,  2016, \mn@doi [\apjl]
  {10.3847/2041-8205/830/2/L26}, \href
  {http://adsabs.harvard.edu/abs/2016ApJ...830L..26S} {830, L26}

\bibitem[\protect\citeauthoryear{{Schneider} et~al.,}{{Schneider}
  et~al.}{2013}]{Schneider2013}
{Schneider} M.~D.,  et~al., 2013, \mn@doi [\mnras] {10.1093/mnras/stt855},
  \href {http://adsabs.harvard.edu/abs/2013MNRAS.433.2727S} {433, 2727}

\bibitem[\protect\citeauthoryear{{Springel} et~al.,}{{Springel}
  et~al.}{2008}]{Springel2008}
{Springel} V.,  et~al., 2008, \mn@doi [\mnras]
  {10.1111/j.1365-2966.2008.14066.x}, \href
  {http://adsabs.harvard.edu/abs/2008MNRAS.391.1685S} {391, 1685}

\bibitem[\protect\citeauthoryear{{Stan Development Team}}{{Stan Development
  Team}}{2016b}]{pystan}
{Stan Development Team} 2016b, Stan Development Team. 2016. PyStan: the Python
  interface to Stan, Version 2.14.0.0., \url{http://mc-stan.org}

\bibitem[\protect\citeauthoryear{{Stan Development Team}}{{Stan Development
  Team}}{2016a}]{stan}
{Stan Development Team} 2016a, Stan Modeling Language Users Guide and Reference
  Manual, Version 2.14.0., \url{http://mc-stan.org}

\bibitem[\protect\citeauthoryear{{Tenneti}, {Singh}, {Mandelbaum}, {Matteo},
  {Feng}  \& {Khandai}}{{Tenneti} et~al.}{2015}]{Tenneti2015a}
{Tenneti} A.,  {Singh} S.,  {Mandelbaum} R.,  {Matteo} T.~D.,  {Feng} Y.,
  {Khandai} N.,  2015, \mn@doi [\mnras] {10.1093/mnras/stv272}, \href
  {http://adsabs.harvard.edu/abs/2015MNRAS.448.3522T} {448, 3522}

\bibitem[\protect\citeauthoryear{{Torrealba}, {Koposov}, {Belokurov}  \&
  {Irwin}}{{Torrealba} et~al.}{2016a}]{Torrealba2016a}
{Torrealba} G.,  {Koposov} S.~E.,  {Belokurov} V.,   {Irwin} M.,  2016a,
  \mn@doi [\mnras] {10.1093/mnras/stw733}, \href
  {http://adsabs.harvard.edu/abs/2016MNRAS.459.2370T} {459, 2370}

\bibitem[\protect\citeauthoryear{{Torrealba} et~al.,}{{Torrealba}
  et~al.}{2016b}]{Torrealba2016b}
{Torrealba} G.,  et~al., 2016b, \mn@doi [\mnras] {10.1093/mnras/stw2051}, \href
  {http://adsabs.harvard.edu/abs/2016MNRAS.463..712T} {463, 712}

\bibitem[\protect\citeauthoryear{{Vera-Ciro}, {Sales}, {Helmi}  \&
  {Navarro}}{{Vera-Ciro} et~al.}{2014}]{VeraCiro2014}
{Vera-Ciro} C.~A.,  {Sales} L.~V.,  {Helmi} A.,   {Navarro} J.~F.,  2014,
  \mn@doi [\mnras] {10.1093/mnras/stu153}, \href
  {http://adsabs.harvard.edu/abs/2014MNRAS.439.2863V} {439, 2863}

\bibitem[\protect\citeauthoryear{{Walker}, {McGaugh}, {Mateo}, {Olszewski}  \&
  {Kuzio de Naray}}{{Walker} et~al.}{2010}]{Walker2010}
{Walker} M.~G.,  {McGaugh} S.~S.,  {Mateo} M.,  {Olszewski} E.~W.,   {Kuzio de
  Naray} R.,  2010, \mn@doi [\apjl] {10.1088/2041-8205/717/2/L87}, \href
  {http://adsabs.harvard.edu/abs/2010ApJ...717L..87W} {717, L87}

\bibitem[\protect\citeauthoryear{{Walker} et~al.,}{{Walker}
  et~al.}{2016}]{Walker2016}
{Walker} M.~G.,  et~al., 2016, \mn@doi [\apj] {10.3847/0004-637X/819/1/53},
  \href {http://adsabs.harvard.edu/abs/2016ApJ...819...53W} {819, 53}

\makeatother
\end{thebibliography}
